\documentclass[prl,aps,twocolumn,preprintnumbers, showpacs, nofootinbib,superscriptaddress,notitlepage]{revtex4-1}
\usepackage{mathrsfs}
\usepackage{amsfonts}
\usepackage{amsmath}
\usepackage{slashed}
\usepackage{array}
\usepackage{verbatim}
\usepackage{epsfig}
\usepackage{graphicx}
\usepackage{color}
\usepackage[dvipsnames]{xcolor}
\usepackage{multirow}
\usepackage{booktabs}
\usepackage{mathrsfs}
\usepackage[normalem]{ulem}

\newcommand{\beq}{\begin{eqnarray}}
\newcommand{\eeq}{\end{eqnarray}}

\begin{document}

\title{Transverse-Momentum-Dependent Wave Functions of Pion from Lattice QCD}

\collaboration{\bf{Lattice Parton Collaboration ($\rm {\bf LPC}$)}}

\author{
\includegraphics[scale=0.10]{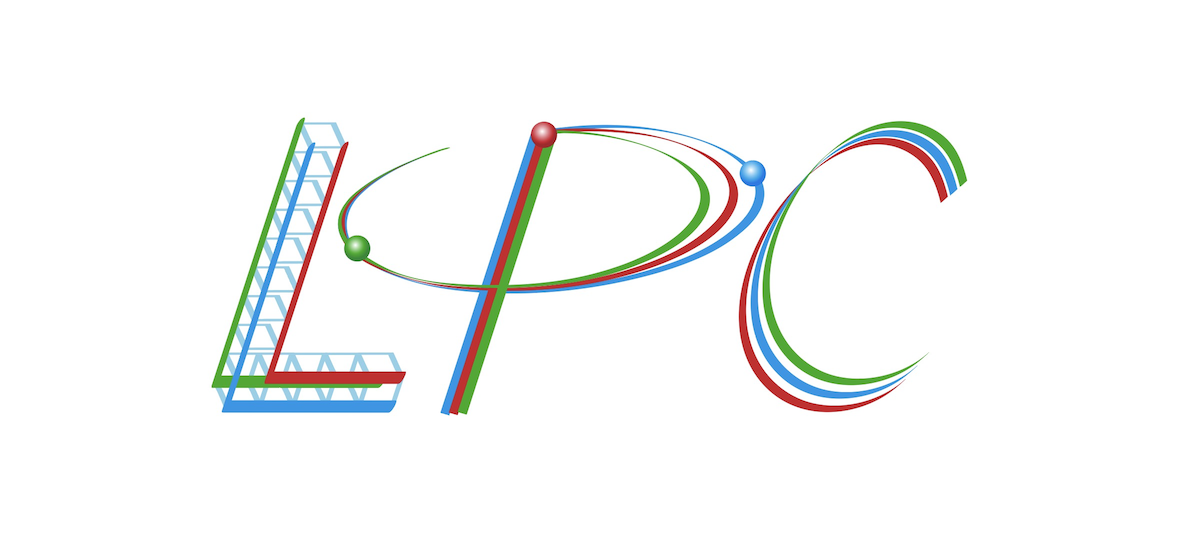}\\
Min-Huan Chu}
\affiliation{INPAC, Shanghai Key Laboratory for Particle Physics and Cosmology, Key Laboratory for Particle Astrophysics and Cosmology (MOE), School of Physics and Astronomy, Shanghai Jiao Tong University, Shanghai 200240, China}
\affiliation{Yang Yuanqing Scientiﬁc Computering Center, Tsung-Dao Lee Institute, Shanghai Jiao Tong University, Shanghai 200240, China}

\author{Jin-Chen He}
\affiliation{INPAC, Shanghai Key Laboratory for Particle Physics and Cosmology, Key Laboratory for Particle Astrophysics and Cosmology (MOE), School of Physics and Astronomy, Shanghai Jiao Tong University, Shanghai 200240, China}
\affiliation{Department of Physics, University of Maryland, College Park, MD 20742, USA}

\author{Jun Hua}
\email{Corresponding author: junhua@scnu.edu.cn}
\affiliation{Guangdong Provincial Key Laboratory of Nuclear Science, Institute of Quantum Matter, South China Normal University, Guangzhou 510006, China}
\affiliation{Guangdong-Hong Kong Joint Laboratory of Quantum Matter, Southern Nuclear Science Computing Center, South China Normal University, Guangzhou 510006, China}

\author{Jian Liang}
\affiliation{Guangdong Provincial Key Laboratory of Nuclear Science, Institute of Quantum Matter, South China Normal University, Guangzhou 510006, China}
\affiliation{Guangdong-Hong Kong Joint Laboratory of Quantum Matter, Southern Nuclear Science Computing Center, South China Normal University, Guangzhou 510006, China}

\author{Xiangdong Ji}
\affiliation{Department of Physics, University of Maryland, College 
Park, MD 20742, USA}

\author{Andreas Sch\"afer}
\affiliation{Institut f\"ur Theoretische Physik, Universit\"at Regensburg, D-93040 Regensburg, Germany}

\author{Hai-Tao Shu}
\email{Corresponding author: hai-tao.shu@ur.de}
\affiliation{Institut f\"ur Theoretische Physik, Universit\"at Regensburg, D-93040 Regensburg, Germany}

\author{Yushan Su}
\affiliation{Department of Physics, University of Maryland, College Park, MD 20742, USA}

\author{Ji-Hao Wang}
\affiliation{CAS Key Laboratory of Theoretical Physics, Institute of Theoretical Physics, Chinese Academy of Sciences, Beijing 100190, China}
\affiliation{School of Fundamental Physics and Mathematical Sciences, Hangzhou Institute for Advanced Study, UCAS, Hangzhou 310024, China}

\author{Wei Wang}
\affiliation{INPAC, Shanghai Key Laboratory for Particle Physics and Cosmology, Key Laboratory for Particle Astrophysics and Cosmology (MOE), School of Physics and Astronomy, Shanghai Jiao Tong University, Shanghai 200240, China}
\affiliation{Southern Center for Nuclear-Science Theory (SCNT), Institute of Modern Physics, Chinese Academy of Sciences, Huizhou 516000, Guangdong Province, China}

\author{Yi-Bo Yang}
\affiliation{CAS Key Laboratory of Theoretical Physics, Institute of Theoretical Physics, Chinese Academy of Sciences, Beijing 100190, China}
\affiliation{School of Fundamental Physics and Mathematical Sciences, Hangzhou Institute for Advanced Study, UCAS, Hangzhou 310024, China}
\affiliation{International Centre for Theoretical Physics Asia-Pacific, Beijing/Hangzhou, China}
\affiliation{School of Physical Sciences, University of Chinese Academy of Sciences,
Beijing 100049, China}

\author{Jun Zeng}
\affiliation{INPAC, Shanghai Key Laboratory for Particle Physics and Cosmology, Key Laboratory for Particle Astrophysics and Cosmology (MOE), School of Physics and Astronomy, Shanghai Jiao Tong University, Shanghai 200240, China}

\author{Jian-Hui Zhang}
\affiliation{School of Science and Engineering, The Chinese University of Hong Kong, Shenzhen 518172, China}
\affiliation{Center of Advanced Quantum Studies, Department of Physics, Beijing Normal University, Beijing 100875, China}

\author{Qi-An Zhang}
\affiliation{School of Physics, Beihang University, Beijing 102206, China}

\begin{abstract}
We present a first lattice QCD calculation of the transverse-momentum-dependent wave functions (TMDWFs) of the pion using large-momentum effective theory. Numerical simulations are based on one ensemble with 2+1+1 flavors of highly improved staggered quarks action with lattice spacing $a=0.121$~fm from the MILC Collaboration, and one with 2 +1 flavor clover fermions and tree-level Symanzik gauge action generated by the CLS Collaboration with $a=0.098$~fm.  As a key ingredient, the soft function
is first obtained by incorporating the one-loop perturbative contributions and a proper normalization. Based on this and the equal-time quasi-TMDWFs simulated on the lattice, we extract the light-cone TMDWFs. The results are comparable between the two lattice ensembles and a comparison with phenomenological parametrization is made. Our studies provide a first attempt of $ab$ $initio$ calculation of TMDWFs which will eventually  lead to crucial theory inputs for making predictions for exclusive processes under QCD factorization.
\end{abstract}

\maketitle
\textit{Introduction:} 
The light-front wave functions (LFWFs) are an important quantity for hadrons in particle physics. They characterize the nonperturbative structure of hadrons, and enter the prediction of a wide variety of measurable observables through quantum chromodynamics (QCD) factorization. While searching for new physics beyond the standard model (SM) requires a dedicated study of high-energy processes at colliders, this goal can partially be achieved by investigating low-energy processes, among which the flavor-changing-neutral-current (FCNC) in a heavy quark system is an ideal probe~\cite{Buchalla:1995vs}. A key input of calculating  the  SM contributions to the FCNC are LFWFs, including the collinear distribution amplitudes (LCDAs) and the transverse-momentum-dependent wave functions (TMDWFs). LFWFs in fact play an essential role in light-front quantization. In particular, the parton distribution functions can be expressed in terms of the square of the TMDWFs~\cite{Brodsky:2001wx,Burkardt:2002uc}. The TMDWFs are characterized by physics at distance scale of a fermi or equivalently momentum scale of a few hundred MeV, which are similar to the confinement scale. Experimental mappings and theoretical computations of these distributions may help to reveal the nature of non-perturbative phenomena such as confinement and chiral symmetry breaking in QCD. 

Although TMDWFs describe important aspects of the three-dimensional structure of hadrons, 
they have never been studied in the literature from the first principles of QCD with systematic approximation. Similar with transverse momentum dependent parton distribution functions (TMDPDFs), it is nontrivial to present a rigorous definition of TMDWFs~\cite{Ma:2004ay}. A key difficulty resides in the rapidity divergences that show up in regularizing the soft contributions from a collinear constituent~\cite{Collins:1981uw}. Therefore, most applications of TMD factorization to hard exclusive processes have  adopted phenomenological models to parametrize the TMDWFs~\cite{Keum:2000ph,Lu:2000em,Ali:2007ff}, which inevitably introduce uncontrollable systematic uncertainties and challenge the precision tests of the SM and probes for new physics. 

Large-momentum effective theory (LaMET)~\cite{Ji:2013dva,Ji:2014hxa} develops a novel way to extract parton physics from the lattice QCD calculations through expansion in large hardon momentum (see~\cite{Ji:2020ect} for a review and many references therein). For TMDWFs, the calculation requires the knowledge on the so-called soft function, which incorporates the effects of soft gluon radiation from colored collinear particles from two opposite light-like directions~\cite{Collins:1981uk,Collins:1988ig}. It was recently discovered that the soft function can be determined by calculating a large-momentum-transfer form factor of a light meson and  quasi TMDWFs on the lattice~\cite{Ji:2019ewn,Ji:2019sxk}, which  removes the obstacle in calculating the TMDWFs from the lattice QCD~\cite{Ji:2021znw, Ji:2020ect}. 
 
In this Letter, we report a first lattice QCD calculation of the pion TMDWFs using LaMET. The calculation is performed on two lattice ensembles with three hadron momenta up to $2.63$~GeV.  We obtain the soft function by incorporating the one-loop perturbative contributions and a proper normalization. 
Based on this, we present first results for the physical TMDWFs. Comparable behaviors between the two lattice ensembles are found and a comparison with the phenomenological model is shown.

\textit{Theoretical Framework:}
The TMDWF $\Psi^{\pm}\left(x, b_{\perp}, \mu, \zeta \right)$ provides the momentum distribution between the quark and antiquark in its leading Fock state. The superscript ``$\pm$" denotes that in $\Psi^{\pm}$ Wilson lines  will approach the positive and negative infinity along the lightcone direction. $x$ denotes the momentum fraction in longitudinal direction, and $b_\perp$ is the Fourier conjugate of transverse momentum. In addition, TMDWFs also depend on the renormalization scale $\mu$ and the rapidity scale $\zeta$.

LaMET allows to access the TMDWF ${\Psi}^{\pm}$ by simulating an equal-time quasi-TMDWF $\tilde{\Psi}^{\pm}$ defined in Euclidean space. The relation between them follows the factorization formula~\cite{Ji:2019sxk,Ji:2021znw}:
\begin{equation}
\begin{aligned}
&\tilde{\Psi}^{\pm}\left(x, b_{\perp}, \mu, \zeta^z\right) S_I^{\frac{1}{2}}\left(b_{\perp}, \mu\right)  \\
&=H^{\pm}\left(x,\zeta^z, \mu\right) 
e^{ \left[\frac{1}{2} K\left(b_{\perp}, \mu\right) \ln \frac{\mp \zeta^z +i \epsilon}{\zeta} \right] }
\Psi^{\pm}\left(x, b_{\perp}, \mu, \zeta\right) \\ 
&~~~+\mathcal{O}\left(\Lambda_{\mathrm{QCD}}^2 /({x^2 \zeta^z}), {M^2} / {\left(P^z\right)^2}, {1} / {(b_{\perp}^2 \zeta^z})\right)
\label{eq:factorization}
\end{aligned}
\end{equation}
where $\zeta^z = (2P^z)^2$. $S_I\left(b_{\perp}, \mu\right) $ denotes the intrinsic soft function, $K\left(b_{\perp}, \mu\right) $ is the Collins-Soper kernel and  has been calculated on the lattice in~\cite{Shanahan:2021tst, LPC:2022ibr, Schlemmer:2021aij}. $H^{\pm}\left(x, \zeta^z, \mu\right) $ represents a perturbative matching kernel. At one-loop level it is given by~\cite{Ji:2021znw,Deng:2022gzi}:
\begin{equation}
\begin{aligned}
    &H^{\pm}\left(x, \zeta^z, \mu\right)\\
    &=1+\frac{\alpha_s C_F}{4 \pi}\left(-\frac{5 \pi^2}{6}-4+l_{\pm}+\bar{l}_{\pm}-\frac{1}{2}\left(l_{\pm}^2+\bar{l}_{\pm}^2\right)\right), \label{eq:hard_kernel}
\end{aligned}
\end{equation}
where $l_{\pm}=\ln[(-x^2 \zeta^z\pm i\epsilon)/\mu^2]$ and $\bar{l}_{\pm}=\ln[(-\bar x ^2 \zeta^z\pm i\epsilon)/\mu^2]$. $x$ and $\bar x=1-x$ are the momentum fractions of quark and antiquark. Power corrections in LaMET factorization are generically suppressed by factors $\left(\Lambda_{\mathrm{QCD}}^2 / (x^2\zeta^z), M^2 /\left(P^z\right)^2, 1 / (b_{\perp}^2 \zeta^z)\right)$.

\begin{figure}
\centering
\includegraphics[scale=0.2]{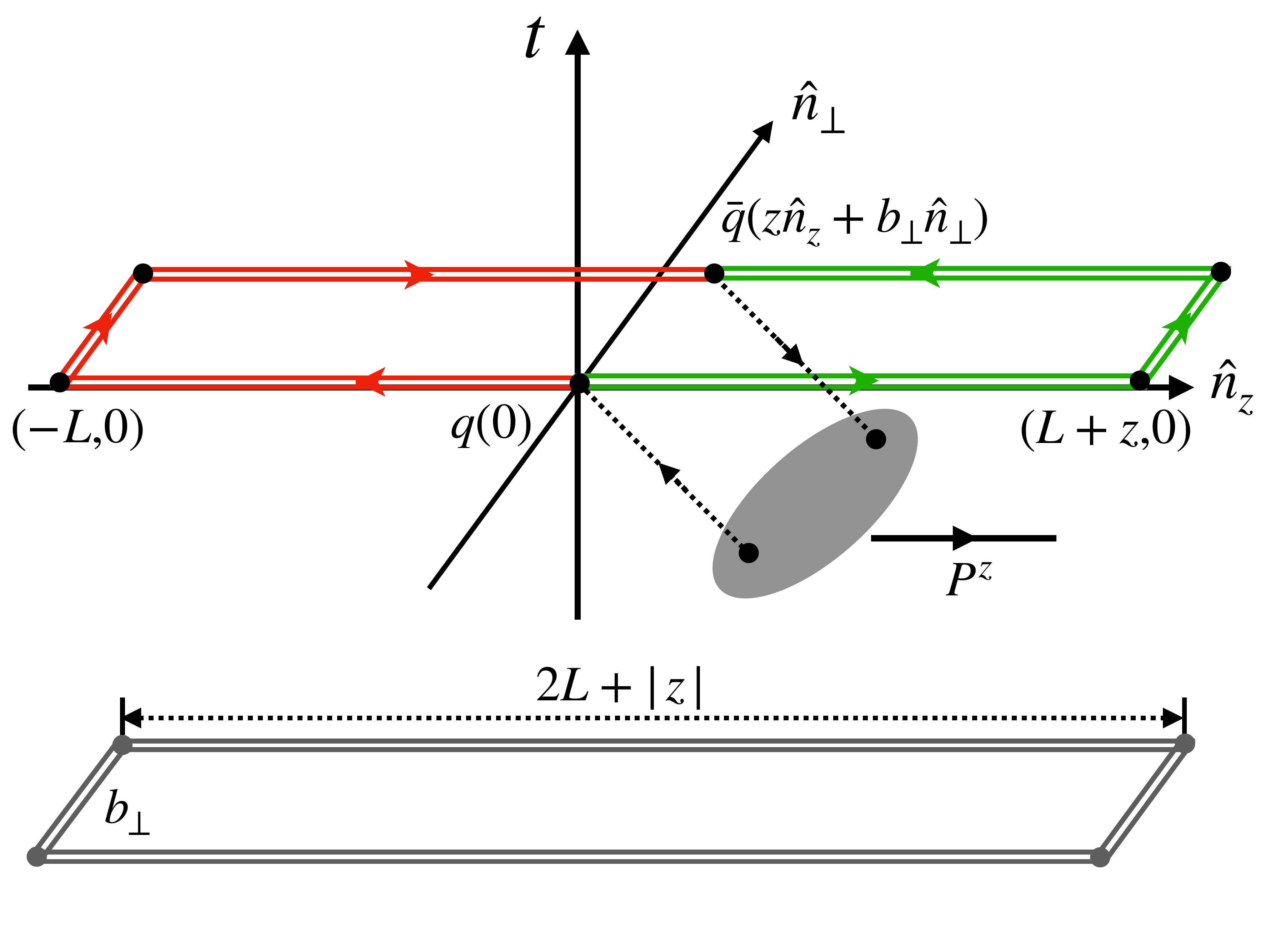}
\caption{Illustration of quasi-TMDWF in coordinate space with a staple-shaped Wilson line inside. The green and red double lines represent the Wilson lines in $\tilde{\Psi}^+(z,b_{\perp},\mu,\zeta^z)$ and $\tilde{\Psi}^-(z,b_{\perp},\mu,\zeta^z)$. A corresponding staple-shaped Wilson loop $Z_E(2L + |z|,b_{\perp},\mu)$ is constructed to cancel the linear and cusp divergences.}
\label{fig:wilson_line}
\end{figure}

In Euclidean lattice, the equal-time quasi-TMDWF 
 in momentum space $\tilde{\Psi}^{\pm}\left(x, b_{\perp}, \mu, \zeta^z\right)$ can be constructed with a   large $P^z$ meson-to-vacuum matrix element of a nonlocal billinear operator for the pseudoscalar meson:
\begin{equation}
\begin{aligned}
&\tilde{\Psi}^{\pm}\left(x, b_{\perp}, \mu,\zeta^z\right) =\lim _{L \rightarrow \infty} \frac{1}{-i f_\pi P^z}\int \frac{d z P^z }{2 \pi} e^{i x z P^z} \\
&~~~~~~~~ \times \frac{\left\langle 0\left|\bar{q}
\left(z \hat{n}_z+b_{\perp} \hat n_\perp\right)
\gamma^t \gamma_5U_{c\pm} q(0)\right| \pi\left(P^z\right)\right\rangle}{\sqrt{Z_E\left(2 L + |z| , b_{\perp}, \mu\right) }Z_O(1/a,\mu)},\label{eq:quasi-TMDWF}
\end{aligned}
\end{equation}
where we choose $\gamma^t\gamma_5$ to project onto the leading-twist TMDWF. The staple-shaped Wilson line between the quark fields $U_{c\pm}$ is required as:
\begin{equation}
\begin{aligned}
U_{c\pm}&= 
U_z^\dagger(z \hat{n}_z+b_{\perp} \hat n_\perp; - \bar{L}_\pm) U_\perp(\bar{L}_\pm\hat{n}_z + z\hat{n}_z;b_\perp)\\
& \times U_z(0\hat{n}_z; \bar{L}_\pm + z),
\label{eq:staple_link}
\end{aligned}
\end{equation}
where $U_{\mu}(x;l)\equiv U_{\mu}(x,x+l\hat{n}_{\mu})$ and $\bar{L}_\pm \equiv \pm \mathrm{max}(L,L\mp z)$, see Fig.~\ref{eq:factorization}. 
$L$ is the length of path-ordered Euclidean Wilson lines along the $z$-direction which will take the $L\to \infty$ limit. But in lattice calculation, one can adopt a sufficiently large $L$. Based on the discussion in~\cite{LPC:2022ibr}, we adopt $L\simeq0.7$~fm in our lattice simulation.

The bare matrix element in the numerator in Eq.~(\ref{eq:quasi-TMDWF}) contains both pinch pole singularity and linear divergence which can be removed by the Wilson loop $Z_E\left(2 L + |z| , b_{\perp}, \mu\right)$~\cite{Ji:2019sxk}. The logarithmic divergences arising from the endpoints of the Wilson line need an additional quark Wilson line vertex renormalization factor $Z_O(1/a,\mu)$. 
A straightforward way to determine $Z_O$ is to evaluate the quotient of the renormalized quasi-TMDWF calculated on the lattice in the small $b_\perp$ region and the quasi-TMDWF perturbatively calculated in $\overline{\mathrm{MS}}$ scheme, as discussed in~\cite{Zhang:2022xuw}. In practice, we adopt $Z_O = \{0.917(2), 0.903(2)\}$ for MILC and CLS ensembles, for details see the Supplemental Material~\cite{supplemental}.


\textit{Lattice simulation:}\
We use one ensemble of the HYP-smeared clover valence fermions action on 2+1+1 flavors of highly improved staggered sea quarks (HISQ)~\cite{Follana:2006rc} generated by MILC~\cite{MILC:2012znn} at the lattice spacing $a=0.121$~fm, and one ensemble of 2+1 flavors clover fermions generated by the CLS Collaboration at $a=0.098$~fm with the unitary valence fermion action. The rest of the simulation setups are collected in  Table.~\ref{Tab:setup}. To improve the signal-to-noise ratio, we adopt hypercubic (HYP) smeared fat links~\cite{Hasenfratz:2001hp} for the staple-shaped gauge link $U_{c\pm}$, and generate the Coulomb gauge fixed wall source propagators $S_w$ to build correlation functions. 
To access the large-momentum limit, we employ three different hadron momenta $P^z = 2\pi/n_s \times \{4,5,6\} = \{1.72,2.15,2.58\}\ \mathrm{GeV}$ for the MILC ensemble and $P^z = 2\pi/n_s \times \{6,8,10\} = \{1.58, 2.11, 2.64\}\ \mathrm{GeV}$ for the CLS ensemble.

\begin{table}
\centering
\caption{The numerical simulation setup. On each ensemble, we put 8/4 source slices in time direction.}
\label{Tab:setup}
\begin{tabular}{cclccccccc}
\hline
\hline
Ensemble ~~& $a$(fm) & \ \!$n_s^3\times\  n_t$   & $m^{sea}_{\pi}$  & $m^{val}_{\pi}$ & Measure \\
\hline
a12m310  ~~& 0.121  ~~& $24^3\times$~ \!64  ~~& 310 MeV         ~~& 670 MeV  &  ~~ 1053$\times$8  ~   \\
\hline
X650  ~~& 0.098  ~~& $48^3\times$~ \!48  ~~& 333 MeV         ~~& 662 MeV  &  ~~ 911$\times$4  ~   \\
\hline
\end{tabular}
\end{table}

To determine the quasi-TMDWF, one can construct the non-local two point correlation function as follows:
\begin{equation}
\begin{aligned}
&C_2^{\pm}(L,z,b_{\perp},t,P^z)\\
&=\sum_{\vec{x}}e^{{\rm i}P^z \vec{x}\cdot \hat{n}_z}\langle S_w^{\dagger}(\vec{x}+z\hat{n}_z+b_{\perp}\hat{n}_{\perp},t)U_{c\pm}S_w(\vec{x},t)\rangle
\end{aligned}
\end{equation}
Due to the limited $L$ in lattice simulation discussed in Eq.~(\ref{eq:staple_link}), we adopt $(z>0)$ for $C_2^+$ and $(z<0)$ for $C_2^-$ in numerical practice, while the remainder can be obtained by isospin symmetry. Such a symmetry behavior in quasi-TMDWF for $\pm z$ have been numerically shown in~\cite{LPC:2022ibr}.

 The ground-state contribution to the quasi-TMDWF can be extracted by the following two-state fit parametrization:
\begin{equation}
\begin{aligned}
&\frac{C^{\pm}_2(L,z,b_{\perp},t,P^z)}{C^{\pm}_2(L,z=0,b_{\perp}=0,t,P^z)}\\
&=\tilde{\Psi}^{\pm, 0}(z,b_{\perp},\zeta^z, L)\frac{1+c_0(z,b_{\perp},P^z,L)e^{-\Delta Et}}{1+c_1e^{-\Delta Et}},\\
\label{eq:c2_fit}
\end{aligned}
\end{equation}
where $\tilde{\Psi}^{\pm, 0}(z,b_{\perp},\zeta^z, L)$ is the bare quasi-TMDWF in coordinate space, while $c_{0,1}$ and $\Delta E$ are free parameters accounting for excited state contamination. In the large $t$ limit, this contamination is suppressed exponentially, which gives the possibility to extract the quasi-TMDWF through a one-state parametrization. Based on the comparison of one- and two-state fits in Supplemental Material~\cite{supplemental}, we find that the one-state fit  gives  a more stable result which will be used in the following analysis.

\textit{Numerical results:}
\begin{figure}
\centering
\includegraphics[scale=0.6]{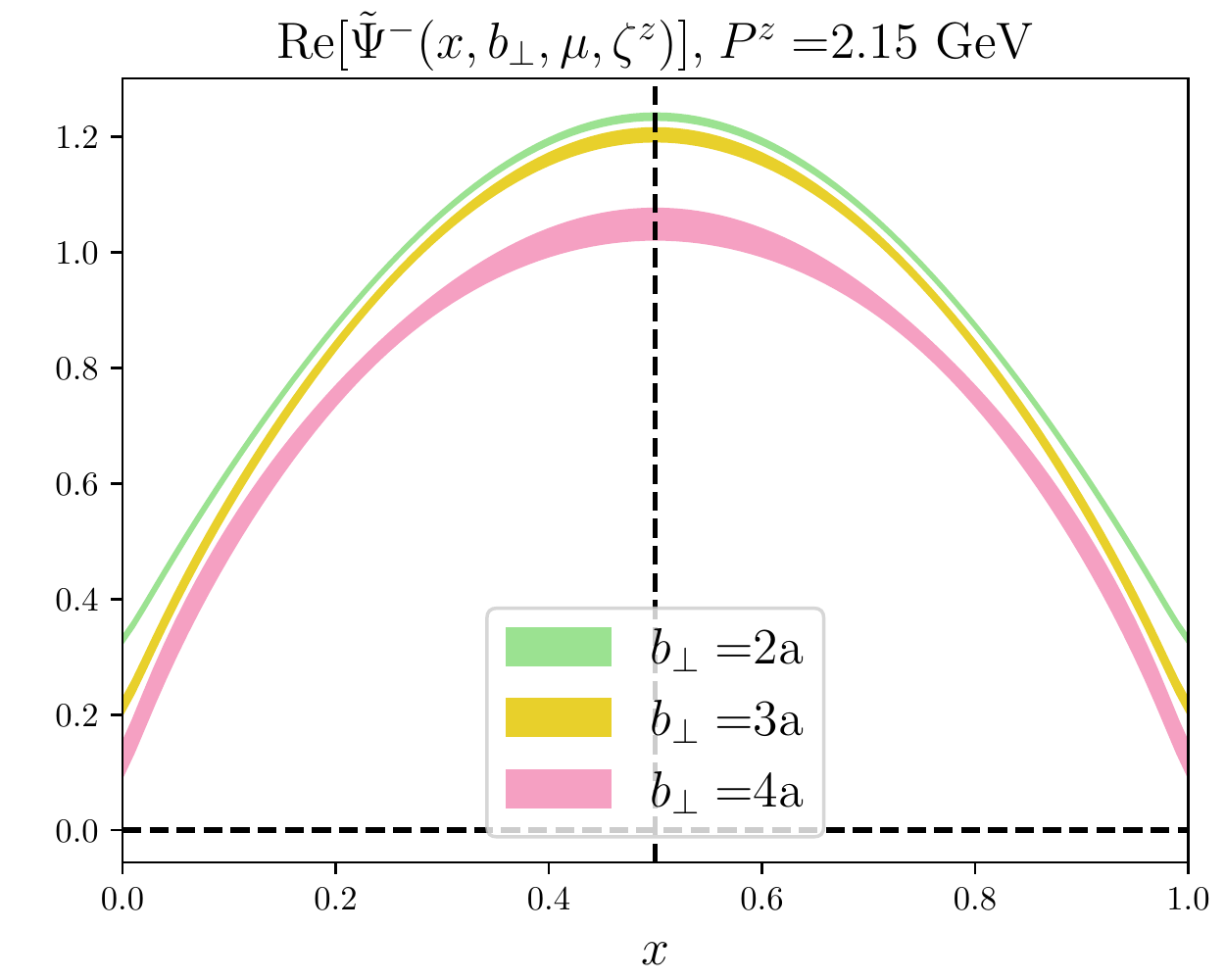}
\includegraphics[scale=0.6]{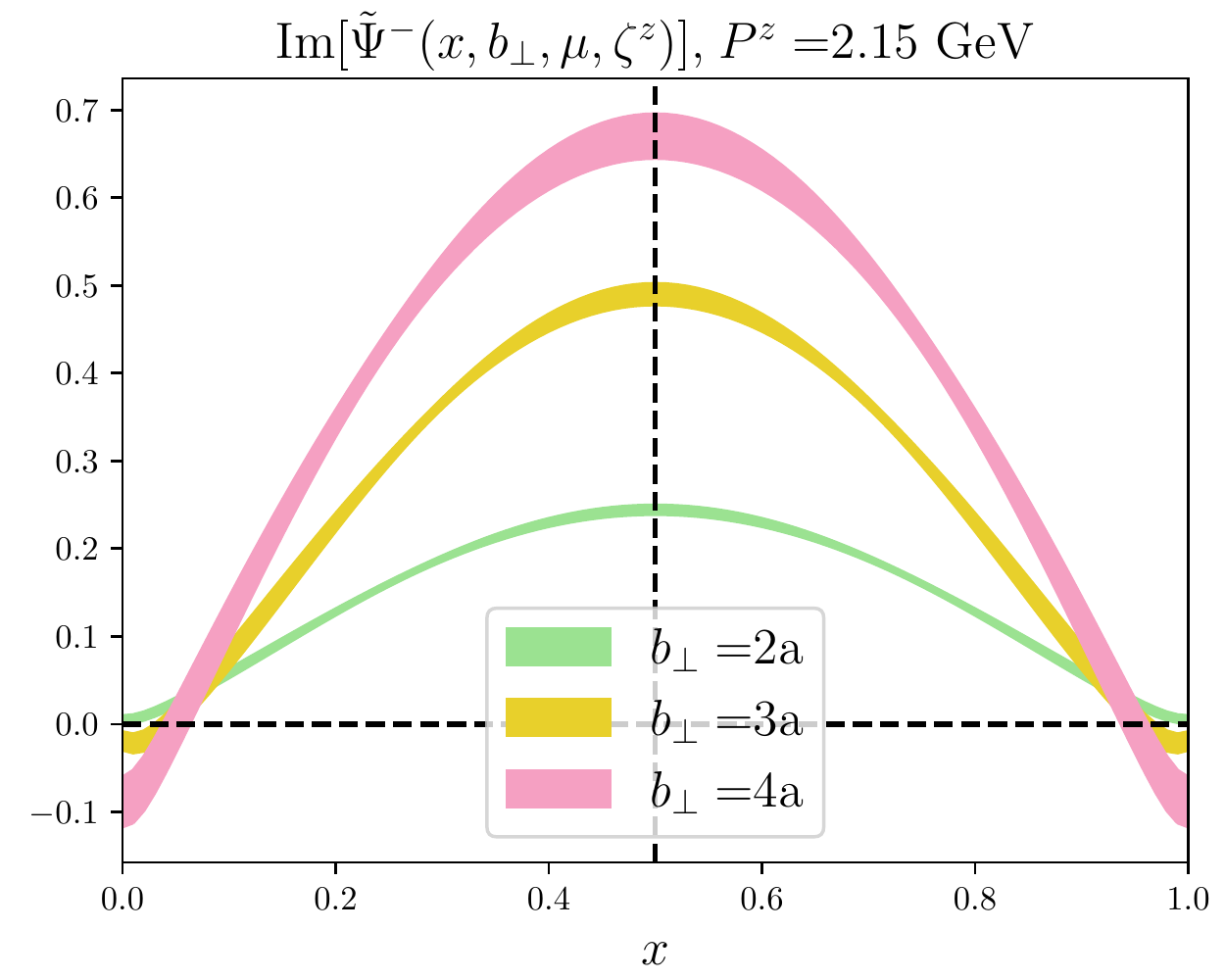}
\caption{The real part (upper panel) and the imaginary part (lower panel) of the quasi-TMDWF in momentum space, with hadron momentum $P^z=2.15$~GeV on MILC ensemble.}
\label{fig:qwf_co}
\end{figure}
After renormalization by Wilson loop $Z_E$ and quark Wilson line vertex correction $Z_O$ referring to Eq.~(\ref{eq:quasi-TMDWF}), the quasi-TMDWF in coordinate space can be obtained straightforwardly. As discussed in a hybrid scheme~\cite{Ji:2020brr}, in the Fourier transformation, a brute-force truncation at finite $z$ will introduce unphysical oscillations. To avoid these oscillations, we adopt an analytical extrapolation at large light front (LF) distance ($\lambda = zP^z$) for quasi-TMDWF in coordinate space:
\begin{equation}
\begin{aligned}
&\tilde{\Psi}(z,b_{\perp},\mu,\zeta^z)= f(b_\perp)\left[\frac{k_1}{(-i\lambda)^d}+e^{i\lambda}\frac{k_2}{(i\lambda)^d}\right]e^{-\frac{\lambda}{\lambda_0}},\label{eq:extrapolation}
\end{aligned}
\end{equation}
where $k_{1,2}, d$ are free parameters, $\lambda_0$ denotes a large distance parameter~\cite{Ji:2020brr, LatticeParton:2022zqc}, and the complex parameter $f(b_\perp)$ describes the behavior in transverse direction. 
After extrapolation and Fourier transformation we get the results shown in Fig.~\ref{fig:qwf_co} for the real part (upper panel) and the imaginary part (lower panel) of the quasi-TMDWF in momentum space at $P^z=2.15$~GeV on MILC ensemble. For $b_\perp = 1a$, there might be sizable discretization effects, and thus we show only the quasi-TMDWF with $b_\perp = \{2,3,4\}a$. As can be seen from this figure, the real part decreases slowly with the increasing $b_\perp$, while the imaginary part increases rapidly with $b_\perp$. Unlike the one dimensional quasi distribution amplitude in~\cite{LatticeParton:2022zqc}, the quasi-TMDWF has a sizable nonzero imaginary part.

\begin{figure}
\centering
\includegraphics[scale=0.6]{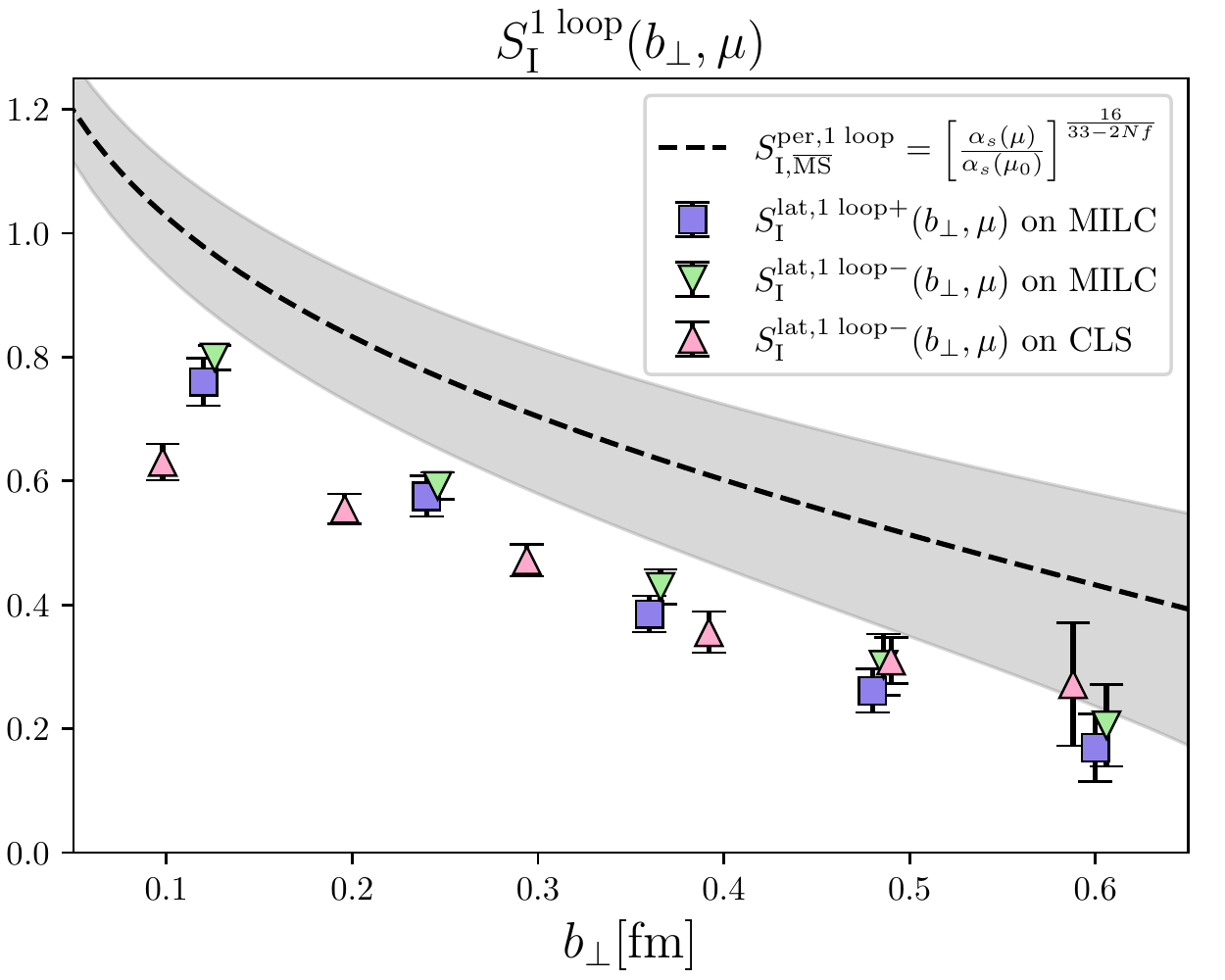}
\caption{The one-loop intrinsic soft function as a function of $b_\perp$. The grey band corresponds to the one-loop perturbative result in the $\overline{\mathrm{MS}}$ scheme and the band is obtained by $\mu_0=1/b^*_{\perp}$ varying in the range $b^*_{\perp}\in\left[1/\sqrt{2},\sqrt{2}\right]b_{\perp}$. 
The label $\pm$ in $S^{\mathrm{lat,1\;loop}\pm}$ represents the lattice results extracted by $\tilde{\Psi}^{\pm}$. }
\label{fig:si_final}
\end{figure}

According to the LaMET factorization  in Eq.~(\ref{eq:factorization}), apart from the quasi-TMDWF, one requires the intrinsic soft function and Collins-Soper (CS) evolution kernel to obtain the TMDWF. In recent years, the CS kernel has been determined on the lattice~\cite{Shanahan:2021tst, LPC:2022ibr, Schlemmer:2021aij}. A recent analysis on MILC ensemble at $0.121$~fm that includes the one-loop perturbative contributions can be found in Ref.~\cite{LPC:2022ibr}, while  on CLS ensemble at $0.098$~fm the result is given in the Supplemental Material~\cite{supplemental}. 

The intrinsic soft function can be determined from the quasi-TMDWF and the form factor of a pseudoscalar meson.  The calculation for tree level intrinsic soft function was performed in~\cite{LatticeParton:2020uhz, Li:2021wvl}. 
Inspired by a detailed theoretical analysis on normalization condition and twist combination of the form factor in~\cite{Deng:2022gzi}, we present the intrinsic soft function in Fig.~\ref{fig:si_final} that is based on  the one-loop matching kernel.  As can be seen from this figure, the intrinsic soft functions extracted by $\tilde{\Psi}^+$ and $\tilde{\Psi}^-$ on MILC ensemble are  consistent with each other, which is in line with the  expectation that the intrinsic soft function is universal. The result obtained from $\tilde{\Psi}^-$ on CLS ensemble is similar but decreases more slowly than the MILC results. A potential reason for this difference might be  the discretization effects, which will be further investigated in future work. Our lattice results have similar $b
_\perp$ dependence as one-loop perturbative result in the  $\overline{\mathrm{MS}}$ scheme~\cite{Ebert:2019okf} in both the small and large $b_\perp$ regions. However, it is necessary to point out that the one-loop perturbative result might be unreliable at large $b_\perp$.

\begin{figure}
\centering
\includegraphics[scale=0.6]{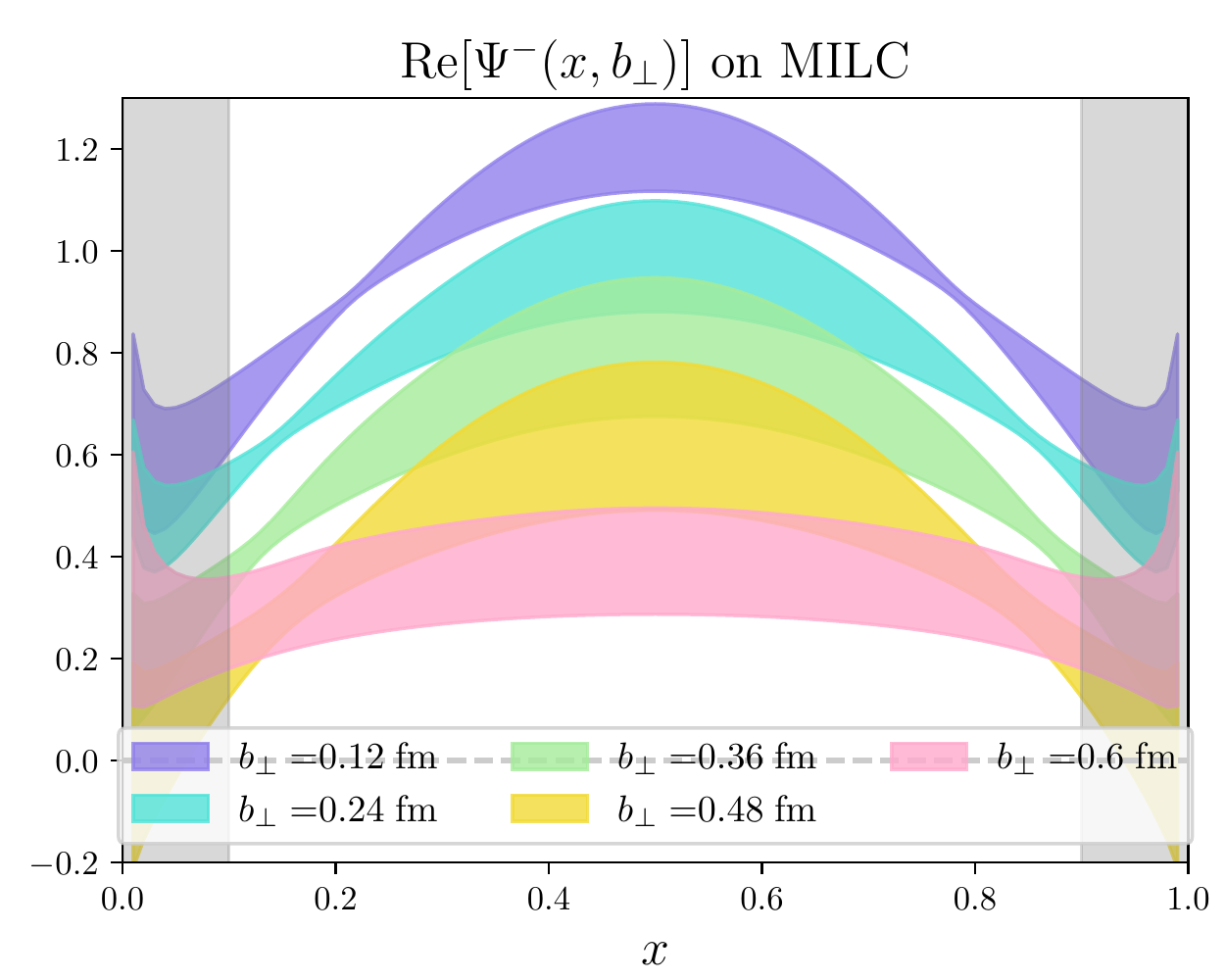}
\includegraphics[scale=0.6]{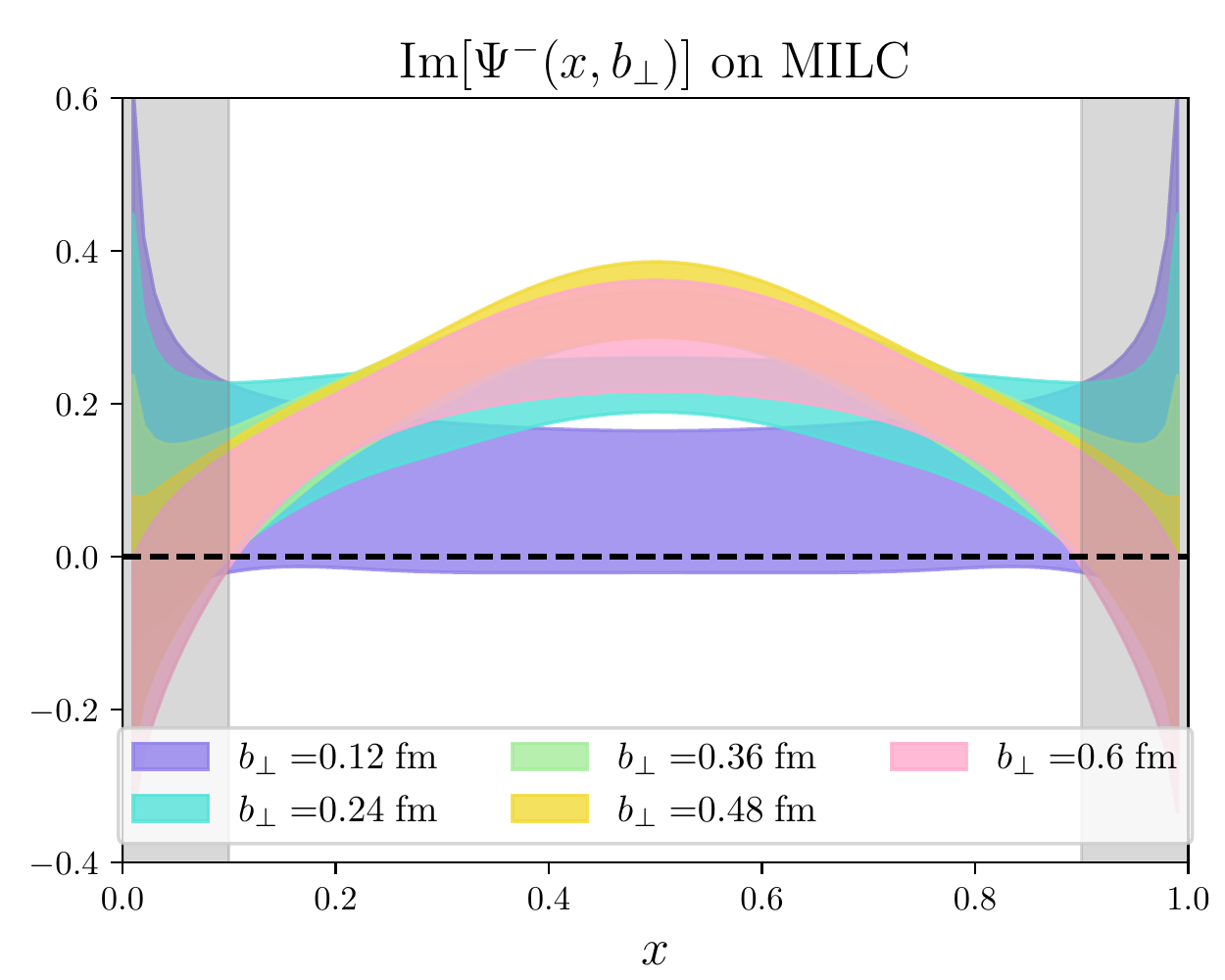}
\caption{The real parts (upper panel) and the imaginary parts (lower panel) of the TMDWF on MILC ensemble. 
The TMDWF results approach the infinite $P^z$ limit with rapidity scale $\zeta=(6\;\mathrm{GeV})^2$ and renormalization scale $\mu=2$~GeV.}
\label{fig:lcwf}
\end{figure}

Together with the quasi-TMDWF, one-loop intrinsic soft function and CS kernel, the TMDWF can be  obtained through a perturbative matching, see Eq.(\ref{eq:factorization}).  In Fig.~\ref{fig:lcwf}, we show the real parts (upper panel) and the imaginary parts (lower panel) of TMDWF $\Psi^-$ calculated on MILC ensemble. Results in this figure contain both statistical and systematic uncertainties, where the systematic ones come from the large $\lambda$ extrapolation and the infinite momentum extrapolation~\cite{supplemental}. The renormalization scale is chosen as $\mu=2\;\mathrm{GeV}$ and the rapidity scale as $\zeta=(2P^+)^2=(6\;\mathrm{GeV})^2$.
As can be seen from the figure, the real part of the TMDWF decreases as $b_\perp$ increases, while the imaginary part first increases and stabilizes for $b_\perp>0.36\;\mathrm{fm}$. 
The imaginary part shows a weaker dependence  on $b_\perp$ than the real part. From  LaMET factorization in Eq.~(\ref{eq:factorization}),  the endpoint region suffers from sizable higher power corrections. With a rough estimation~\cite{Hua:2020gnw} $\lambda \simeq 10$, we conclude that the shaded regions ($x<0.1$ and $x>0.9$) cannot be reliably controlled  in LaMET at present.

\begin{figure}
\centering
\includegraphics[scale=0.65]{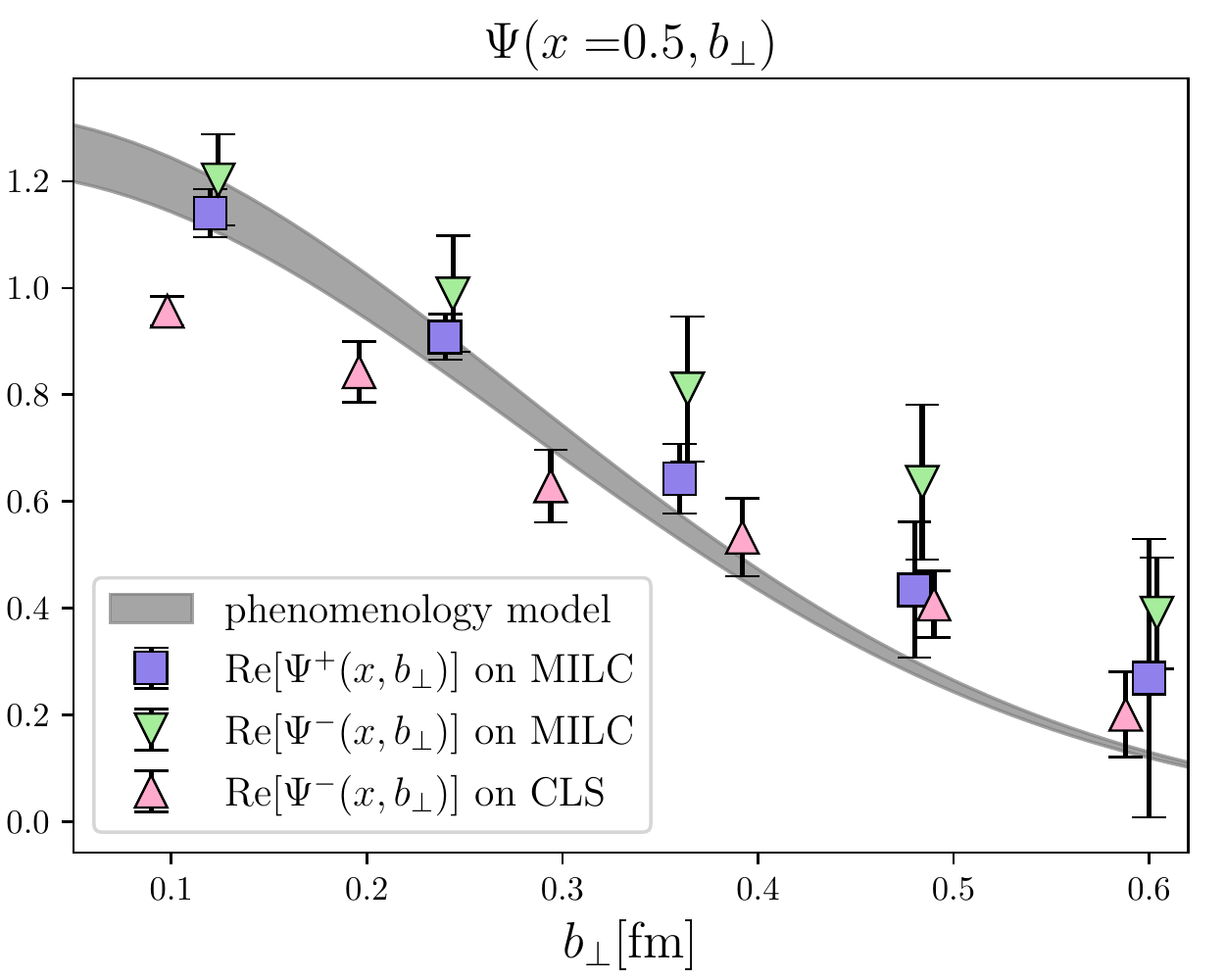}
\caption{Comparison of the transverse momentum distribution in our results with   $\{\zeta,\mu\}=\{(6\;\mathrm{GeV})^2,2\;\mathrm{GeV}\}$ and phenomenological model at $x=0.5$.}
\label{fig:pheno}
\end{figure}

In Fig.~\ref{fig:pheno}, we show a comparison of TMDWFs ${\Psi}^\pm$ at the momentum fraction $x = 0.5$ on MILC ensemble and CLS ensemble with a phenomenological model~\cite{Lu:2007hr}, which factorizes TMDWF into longitudinal and transverse momentum distributions. The TMDWFs decay with increasing $b_\perp$, which is consistent with the phenomenological model. 
However, the phenomenological parametrization only contains the real parts and does not include the difference of Wilson line directions in Eq.~(\ref{eq:staple_link}).
The non-zero imaginary part may introduce additional complexity to  phenomenological applications which has not been discussed in previous analyses.

Our numerical results are based on different discretizations and lattice spacings, thus their difference can be considered as an estimate of the discretization error before further studies at smaller lattice spacings.
Besides, our lattice simulations are performed on pion mass around $670$ MeV, which is far from the physical point. 
Therefore, our results are still subject to large systematic uncertainties, and future calculations with smaller lattice spacings and lighter quark masses can significantly improve them.


\textit{Summary:}\
We present a first lattice calculation of the transverse momentum dependent wave function of the pion. Numerical simulations are conducted on two ensembles by the MILC and CLS collaborations. The linear and logarithmic divergences are cancelled by Wilson loop and quark Wilson line vertex correction. The extrapolation strategy for quasi-TMDWF in coordinate space follows the hybrid scheme.

The final results of TMDWFs extracted from both two ensembles have a consistent $b_\perp$ dependence, with some differences at small $b_\perp$ which would come from discretization errors. 
These results provide a first attempt of $ab$ $initio$ calculation for TMDWFs which will eventually  lead to  crucial theory inputs for making predictions for exclusive processes under QCD factorization.

{\it Acknowledgement}: We thank the CLS Collaboration for sharing the ensembles used to perform this study. We thank Wolfgang S\"oldner for valuable discussions on the X650 ensemble. 
This work is supported in part by Natural Science Foundation of China under grant No. U2032102,  12125503, 12205106, 12175073, 12222503, 12293062, 12147140, 12205180, 12047503, 12005130.  The computations in this paper were run on the Siyuan-1 cluster supported by the Center for High Performance Computing at Shanghai Jiao Tong University, and Advanced Computing East China Sub-center. J.H and J.L are also supported by Guangdong Major Project of Basic and Applied Basic Research No. 2020B0301030008, the Science and Technology Program of Guangzhou No. 2019050001.  Y.B.Y is also supported by the Strategic Priority Research Program of Chinese Academy of Sciences, Grant No. XDB34030303 and XDPB15. J.H.Z. is supported in part by National Natural Science Foundation of China under grant No. 11975051. J.Z. is also supported by the China Postdoctoral Science Foundation under Grant No. 2022M712088. 
A.S., H.T.S, W.W, Y.B.Y and J.H.Z are also supported by a NSFC-DFG joint grant under grant No. 12061131006 and SCHA~458/22.  

\clearpage


\begin{widetext}
\section*{Supplemental Materials}
\end{widetext}

\subsection{Fits for two point functions}
\begin{figure}
\centering
\includegraphics[scale=0.7]{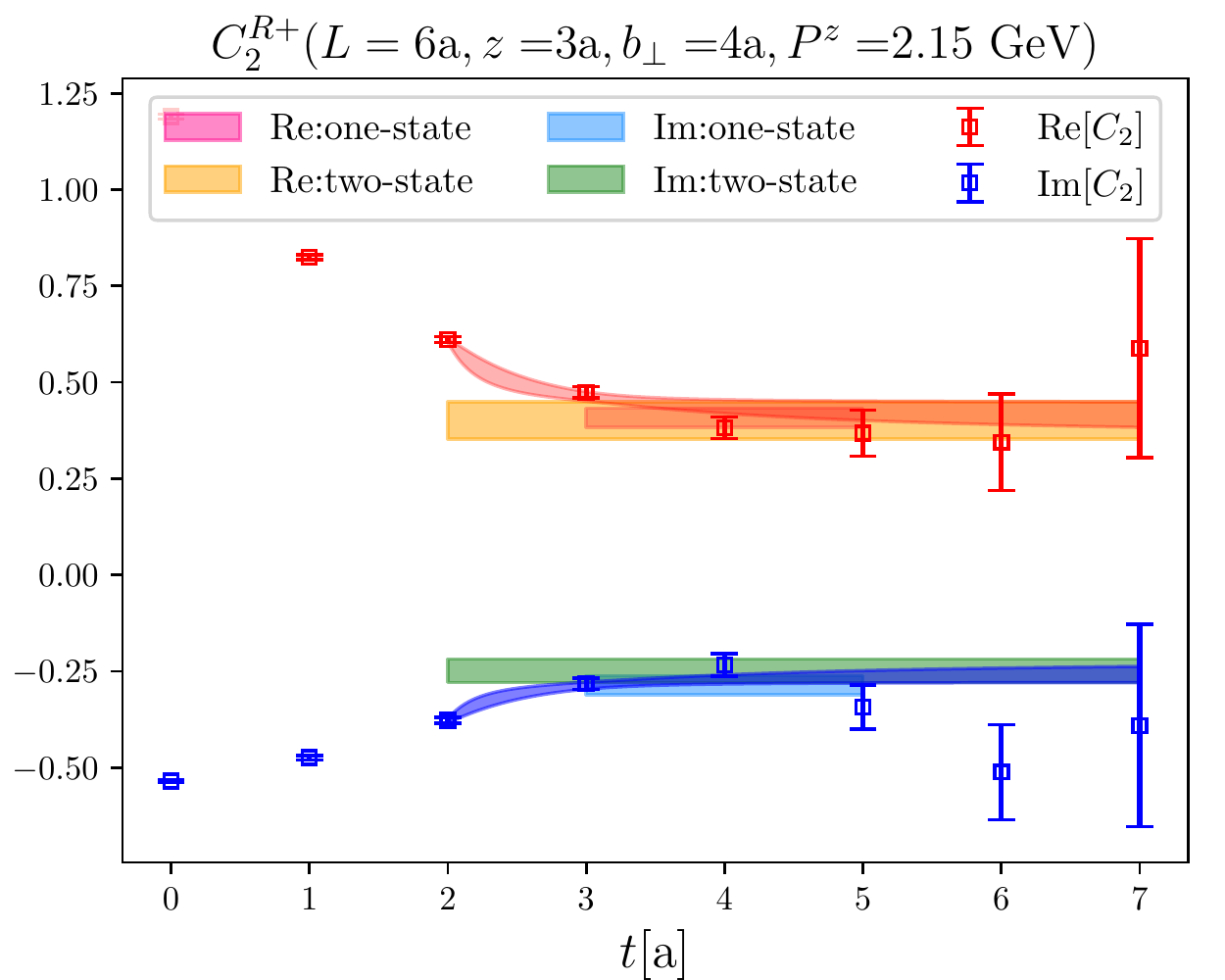}
\caption{One- and two-state fit for $C^{R+}_2(L=6a,z,b_{\perp},P^z,t)$ at $\{L,z,b_{\perp},P^z\}=\{6\mathrm{a},3\mathrm{a},4\mathrm{a},2.15\;\mathrm{GeV}\}$ on MILC ensemble. The red/blue curve corresponds to the two-state fits of the real/imaginary part of $C^{R+}_2$, and the color labeled horizontal bars are the extracted ground-state contributions from one- and two-state fits.}
\label{fig:one-state-fit}
\end{figure}

In lattice simulations, quasi-TMDWFs can be extracted from the two-point correlation functions $C_2^{\pm}$ as shown in Eq.~(\ref{eq:c2_fit}) of the main text. $C_2^{\pm}$ consists of ground-state contributions and  excited-state contaminations, thus one can adopt a two-state fit to separate the ground-state contribution to quasi-TMDWF. In addition, when $t$ becomes large, the excited-state contamination decreases exponentially, so a one-state analysis also allows to extract the ground-state contribution. Fig. \ref{fig:one-state-fit} shows a comparison of one- and two-state fits for the renormalized $C_2^{R+}$ at $\{L,z,b_{\perp}\}=\{6,3,4\}a$ on MILC ensemble, in which they give consistent results. However, the stability of a two-state fit relies strongly on the size of the excited-state contribution, which requires very high precision lattice data. Therefore we employ a one-state fit in the range $t\geq 3a$ to provide reasonable and stable uncertainties in the following analysis.

\subsection{Renormalization}
\begin{figure}
\centering
\includegraphics[scale=0.6]{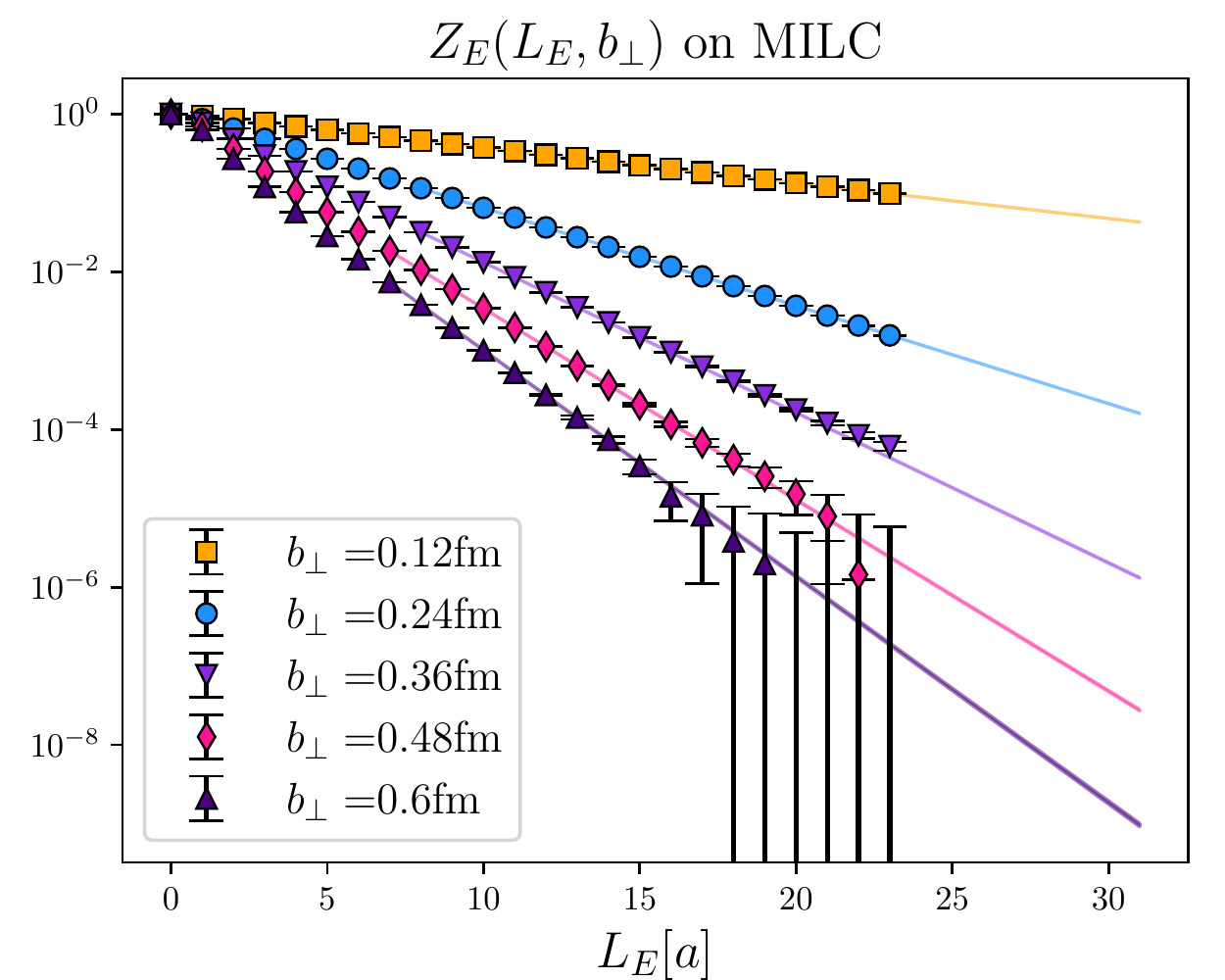}
\caption{Extrapolation of the Wilson loop $Z_E\left(L_E , b_{\perp} \right)$ at $b_{\perp}=\{1,2,3,4,5\}a$ on MILC ensemble.}
\label{fig:wiloop_fit}
\end{figure}

 The linear divergence and pinch pole singularity in quasi-TMDWFs can be removed by the Wilson loop $Z_E\left(L_E , b_{\perp}\right)$, in which $L_E \equiv 2L + |z|$. 
 In lattice simulations, the statistical uncertainty  of $Z_E$ for large $L_E$ and $b_{\perp}$ is out of control as shown in Fig.~\ref{fig:wiloop_fit}.
 Fortunately, the self-energy corrections and gluon exchanging effect introduce linear divergence as exponential form proportional to $L_E$~\cite{Ji:2017oey,LatticePartonCollaborationLPC:2021xdx,Ji:2019ewn}. Therefore one can adopt an extrapolation for the Wilson loop by a two-state fit via the following equation:
\begin{equation}
\begin{aligned}
Z_E(L_E,b_{\perp})=c_0(b_{\perp})e^{-E(b_{\perp})L_E}\left[1+c_1(b_{\perp})e^{-\Delta E(b_{\perp})L_E}\right].
\end{aligned}
\end{equation}
As shown in Fig.~\ref{fig:wiloop_fit}, taking the result on MILC ensemble as an example,  one can see that the extrapolated data is in line with the original data.

\begin{figure}
\centering
\includegraphics[width=0.45\textwidth]{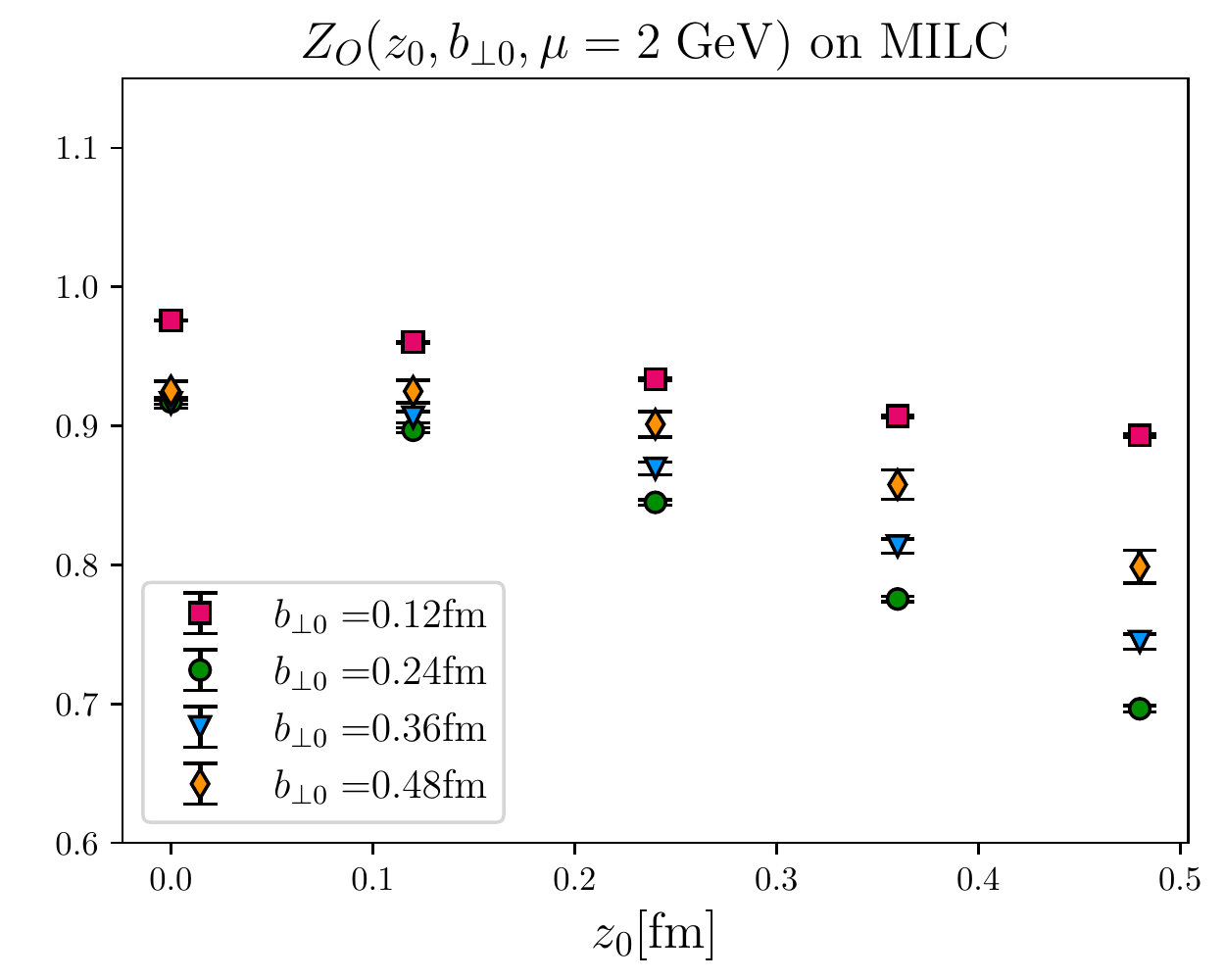}
\includegraphics[width=0.45\textwidth]{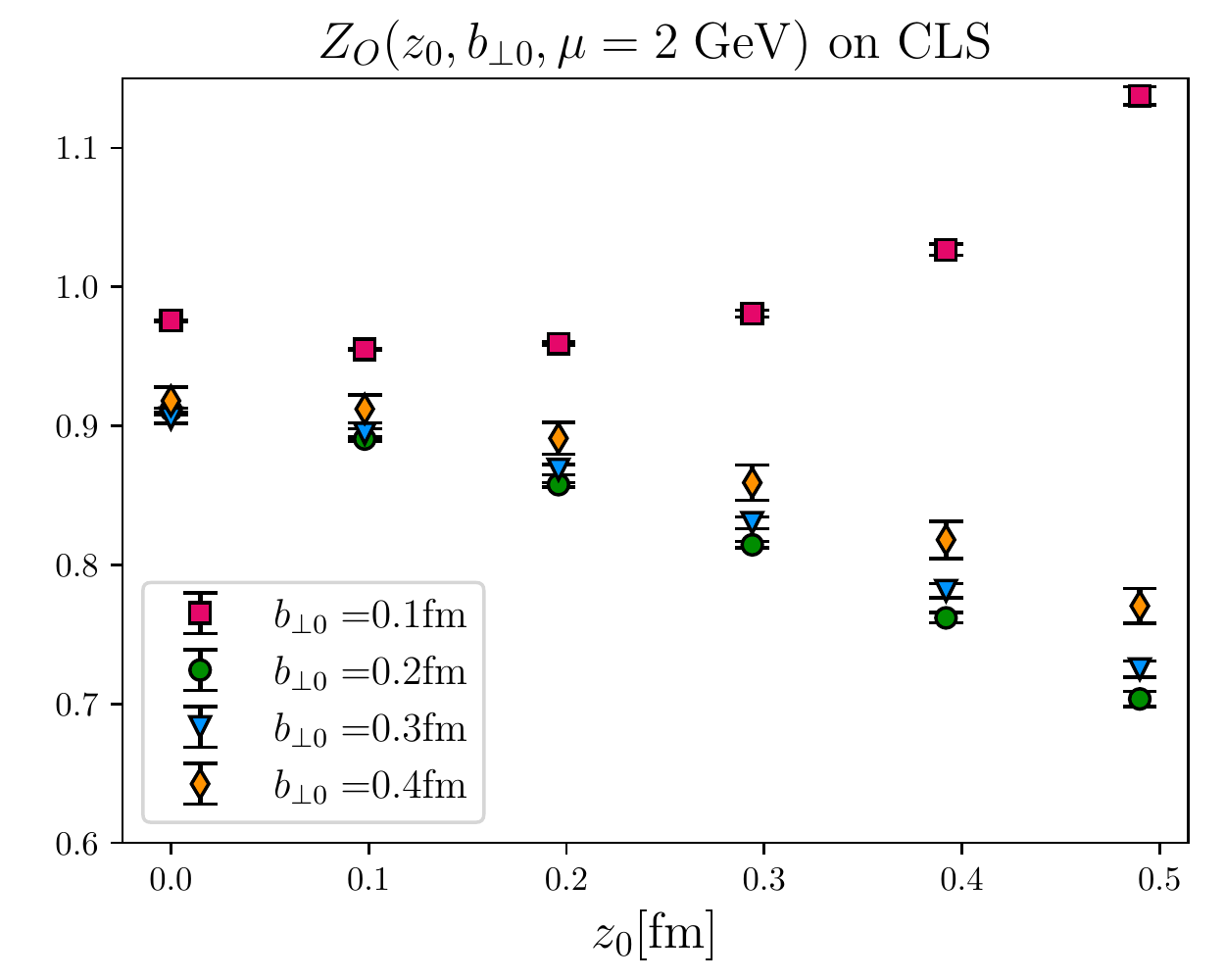}
\caption{The renormalization factor $Z_O$ (see Eq.(\ref{eq:Z_O})) measured at several selected $b_{\perp0}$ on MILC ensemble (upper panel) and on CLS ensemble (lower panel).}
\label{fig:Zo_pic}
\end{figure}

Moreover, to cancel the logarithmic divergence arising from the endpoints of the Wilson lines, we need an additional quark Wilson line vertex renormalization factor $Z_O$. According to Ref.~\cite{Zhang:2022xuw}, this factor $Z_O$  can be computed from the quotient of renormalized quasi-TMDWF in the rest frame calculated on the lattice and perturbatively:
\begin{align}
Z_O(1/a,\mu)=\frac{\tilde{\Psi}^{\pm,0}\left(z_0, b_{\perp 0}, \zeta^z=0, L\right)}{\sqrt{Z_E\left( 2 L + |z_0|, b_{\perp 0}, \mu \right)} \tilde{\psi}^{\overline{\mathrm{MS}}}\left( z_0, b_{\perp 0}, \mu\right)}.\label{eq:Z_O}
\end{align}
$\tilde{\Psi}^{\pm,0}$ denotes the bare quasi-TMDWF in coordinate space. The perturbative quasi-TMDWF $\tilde{\psi}^{\overline{\mathrm{MS}}}$ in the $\overline{\mathrm{MS}}$ scheme in the denominator has been recently calculated in~\cite{Zhang:2022xuw}. 
In our analysis, we adopt a short distance region for $z_0$ and $b_{\perp0}$ matching with perturbative calculation.
 We look for a window of $b_{\perp0}$ where both discretization effects and higher twist contaminations are negligible. In such a window, $Z_O$ should have only a mild dependence on $b_{\perp0}$.
 These dependences are investigated in Fig.~\ref{fig:Zo_pic}. 
 As shown in the figure, $Z_O(z_0=0a, b_{\perp 0}=2a)$ and $Z_O(z_0=0a, b_{\perp 0}=3a)$ reach a $b_{\perp0}$ window for both MILC and CLS ensembles. 
 Such a window is also visible at $z_0=1a$, but becomes invisible as $z_0$ increases. 
 $Z_O(z_0=0a, b_{\perp 0}=0a)$ may suffer from discretization effects since the $b_{\perp0} \rightarrow 0 $ and $a \rightarrow 0$ limits do not commute~\cite{Ji:2020brr}. 
 Thus, $Z_O$ is taken as $\{0.917(2), 0.903(2)\}$ for MILC and CLS ensembles, which is the average of $Z_O(z_0=0a,b_{\perp0}=2a)$ and $Z_O(z_0=0a,b_{\perp0}=3a)$. 

\subsection{Extrapolation for quasi-TMDWF}
\begin{figure}
\centering
\includegraphics[width=0.45\textwidth]{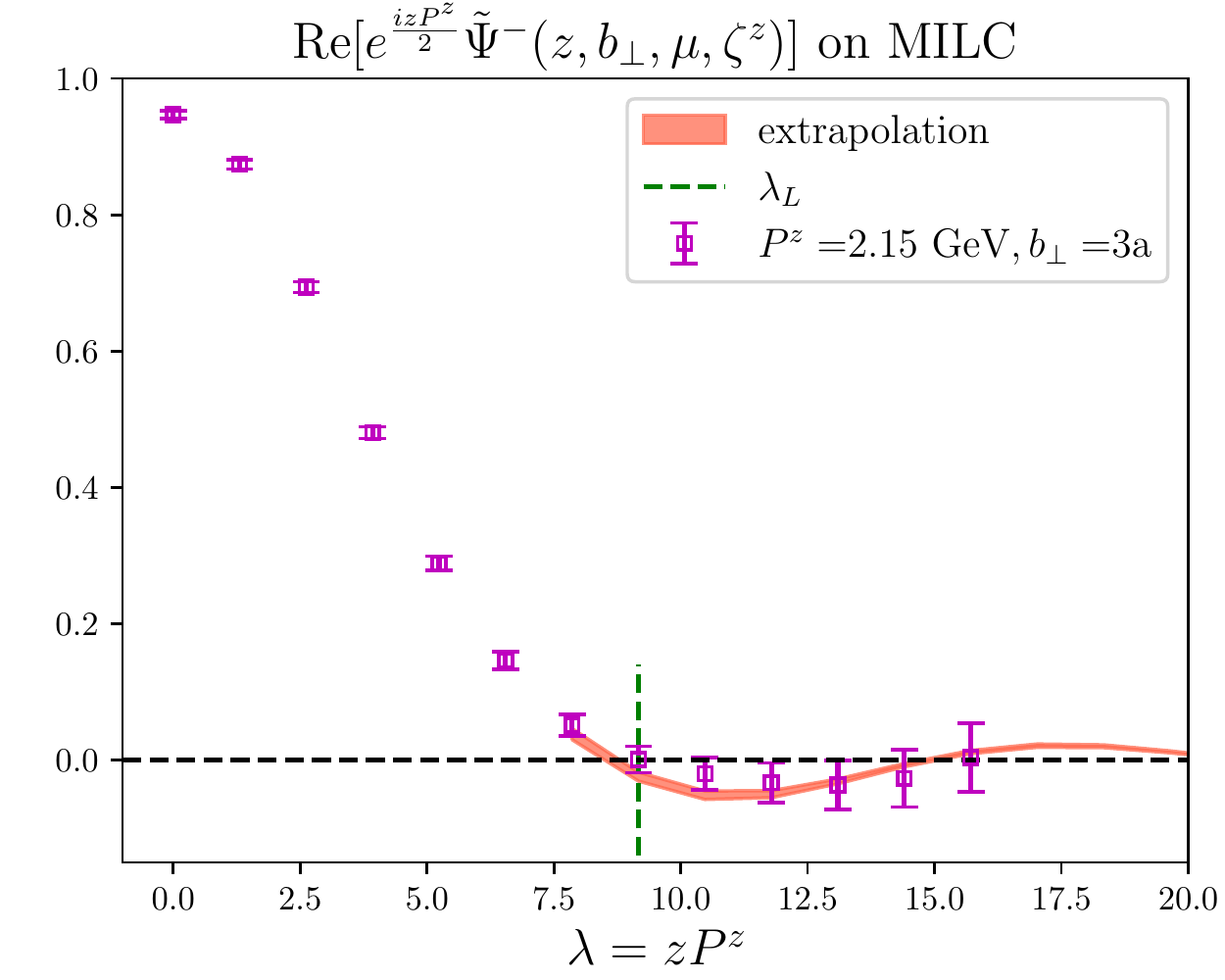}
\caption{Extrapolation of the quasi-TMDWF at large $\lambda$, via a joint fit at different $b_{\perp}$ using Eq.(\ref{eq:extrapolation_ri}). The fits are conducted from the start point of shaded green band to the largest $\lambda$. The data beyond the vertical dashed line are replaced by the fit results. In the panel we show the results on MILC ensemble.}
\label{fig:extrapolation}
\end{figure}

In this Letter, we perform an analytical extrapolation for quasi-TMDWFs in coordinate space to remove unphysical oscillations at large quasi-LF distance ($\lambda=zP^z$) as shown in Eq. (\ref{eq:extrapolation}) of the main text. In lattice simulations, we use the following parametrization for real and imaginary parts:
\begin{align}
&\mathrm{Re}[\tilde{\Psi}^{\pm}(z,b_{\perp},\mu,\zeta^z)]\nonumber\\
&=[k_1(b_\perp)\cos(\frac{\pi d}{2})-k_2(b_\perp)f(b_\perp)\sin(\frac{\pi d}{2})]\frac{e^{-\frac{\lambda}{\lambda_0}}}{\lambda^d},\nonumber\\
&\mathrm{Im}[\tilde{\Psi}^{\pm}(z,b_{\perp},\mu,\zeta^z)]\nonumber\\
&=[k_1(b_\perp)f(b_\perp)\cos(\frac{\pi d}{2})+k_2(b_\perp)\sin(\frac{\pi d}{2})]\frac{e^{-\frac{\lambda}{\lambda_0}}}{\lambda^d}.\label{eq:extrapolation_ri}
\end{align}
The parameter $d$ accounts for the geometric attenuation, and the trigonometric function terms represent the periodic changes. 
Since we have parametrized the transverse-momentum-dependence part by a multiplicative complex number for each $b_\perp$, the parameter $d$ no longer depends on $b_{\perp}$. Therefore we perform a joint fit with $b_{\perp}=\{1,2,3,4,5\}a$ ($a=0.121$~fm) on MILC ensemble and $b_{\perp}=\{1,2,3,4,5,6\}a$ ($a=0.098$~fm) on CLS ensemble. 
As shown in Fig.~\ref{fig:extrapolation}, the uncertainty of the lattice data grows rapidly for large $\lambda$. The extrapolated data matches the original ones and control the uncertainties well.
We adopt $\lambda_{L} = z_{L}P^z \approx 10$ to distinguish the original data ($\lambda < \lambda_{L}$) and the extrapolated ones ($\lambda > \lambda_{L}$). To estimate the modification effects of the extrapolation form, we take additional two extrapolation cases $\lambda_L = (z_{L}\pm 1)P^z$, and treat the average of their differences as an estimate of the systematic uncertainty.

\subsection{Collins-Soper kernel on CLS ensemble}
\begin{figure}
\centering
\includegraphics[scale=0.6]{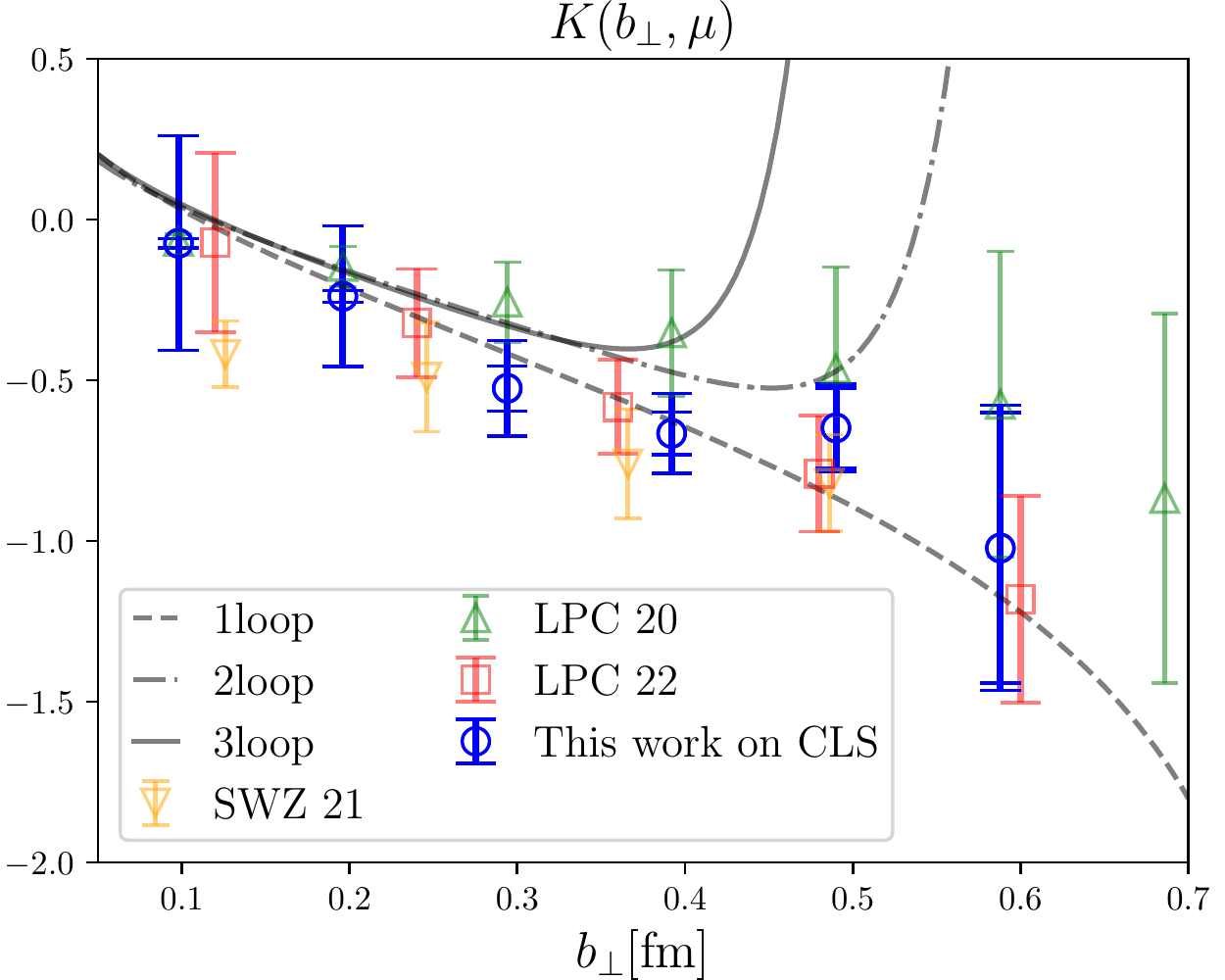}
\caption{The blue data points display our CS kernel results on CLS ensemble, which contain statistical and systematic uncertainties shown as inner error bars and outer ones. SWZ 21~\cite{Shanahan:2021tst}, LPC 20~\cite{LatticeParton:2020uhz} and LPC 22~\cite{LPC:2022ibr} are previous lattice results.}
\label{fig:CS_cls}
\end{figure}

We calculate the Collins-Soper kernel on CLS ensemble in this work inspired by~\cite{LPC:2022ibr}. The result is shown in Fig.~\ref{fig:CS_cls}, in which   previous lattice QCD and perturbative calculations are also included as a comparison. As discussed in~\cite{LPC:2022ibr}, the dominant systematic uncertainty at small $b_\perp$ comes from the imaginary part of the matching kernel.

\subsection{Large momentum extrapolation and TMDWF}
\begin{figure}
\centering
\includegraphics[scale=0.6]{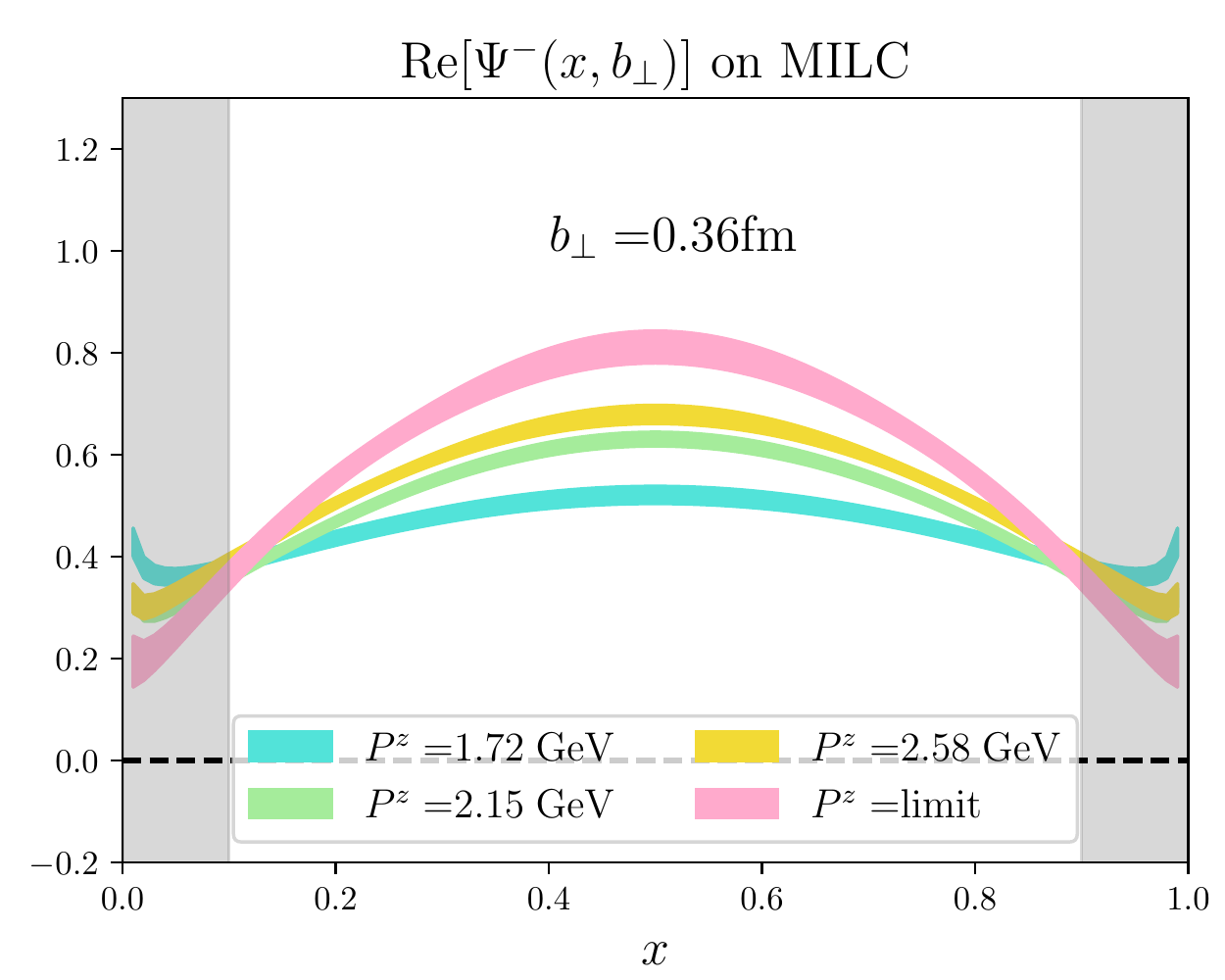}
\includegraphics[scale=0.6]{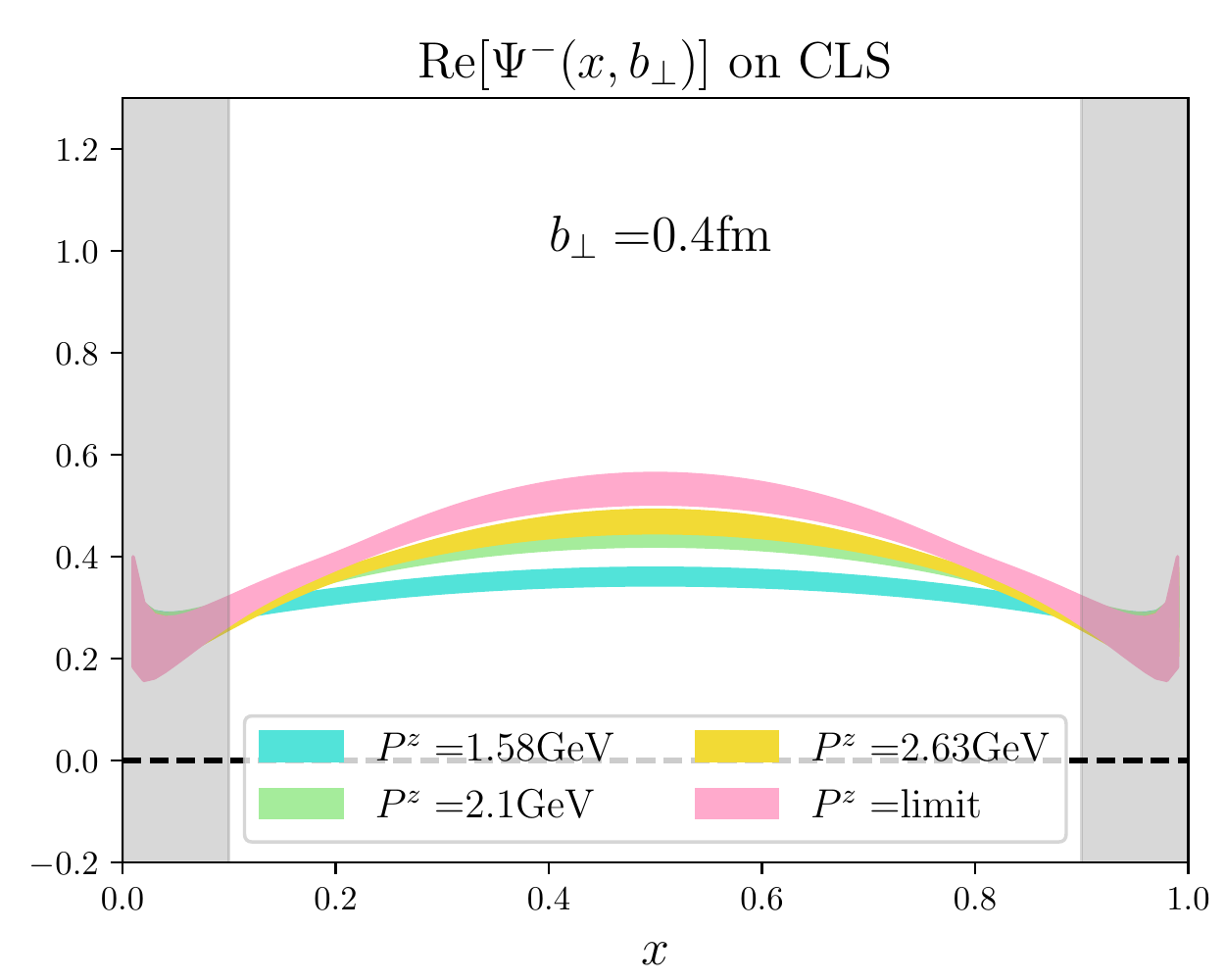}
\caption{The figures display the $P^z$ dependence of TMDWFs at $b_{\perp}=2a$ on MILC ensemble (upper panel) and $b_{\perp}=3a$ on CLS ensemble (lower panel) and their comparison with the infinite $P^z$ limit.}
\label{fig:lcwf_mom_dep}
\end{figure}

In lattice simulations, we adopt an extrapolation to infinite $P^z$ for TMDWFs with the following equation:
\begin{align}
\Psi^{\pm}(P^z)=\Psi^{\pm}(P^z\to\infty)+\frac{A}{(P^z)^2}.
\end{align}
The extrapolations are performed on MILC ensemble with $P^z=\{1.72,2.15,2.58\}$~GeV and on CLS ensemble with $P^z=\{1.58,2.1,2.63\}$~GeV, as shown in Fig.~\ref{fig:lcwf_mom_dep}. The difference between the largest $P^z$ and $P^z\to\infty$ is taken as one of the systematic uncertainties.

Two systematic uncertainties for TMDWFs are considered in our results. As previously illustrated, one is from large $\lambda$ extrapolation of quasi-TMDWFs, the other is from infinite $P^z$ extrapolation. The whole uncertainty is the quadratic summation:
\begin{align}
\sigma_{\mathrm{all}}=\sqrt{\sigma_{\mathrm{stt}}^2+\sigma_{\lambda_L}^2+\sigma^2_{P^z_{\mathrm{lim}}}},
\end{align}
where $\sigma_{\mathrm{stt}}$ represents the statistical uncertainty, $\sigma_{P^z_{\mathrm{lim}}}$ is from infinite momentum extrapolation, and $\sigma_{\lambda_L}$ corresponds to the difference between extrapolation with $\lambda_L=(z_L\pm1)P^z$ and $\lambda_L=z_LP^z$.

In Fig.~12, we show the final results of TMDWFs containing statistical and systematic uncertainties, the TMDWFs $\Psi^{\pm}$ on MILC ensemble (left and central panels) and $\Psi^-$ on CLS ensemble (right panel). 
As one can see from those figures, the real parts of TMDWFs $\Psi^{\pm}$ on both MILC and CLS ensembles decrease with increasing $b_{\perp}$. While the real part of $\Psi^-$ on CLS in small $b_{\perp}$ ($b_{\perp}<0.3$~fm) is smaller compared with $\Psi^-$ on MILC, which might be caused by discretization effects. 
The imaginary parts of TMDWFs in all three cases increase with $b_{\perp}$ and become stable for $b_{\perp}>0.36$~fm. 
In addition, for large $b_{\perp}$ ($b_{\perp}\geq0.48$~fm), the imaginary part of $\Psi^-$ converges, while for $\Psi^+$ it does not. 
The reason for this different behaviour is that in the hard kernels $H^{\pm}(x,\zeta^z,\mu)$ in Eq.(\ref{eq:hard_kernel}) of the main text, the logarithmic term $\ln(-x^2\zeta^2\pm i\epsilon)$ has a different sign of imaginary part for $\Psi^\pm$.

\begin{widetext}

\begin{figure}
\centering
\includegraphics[scale=0.46]{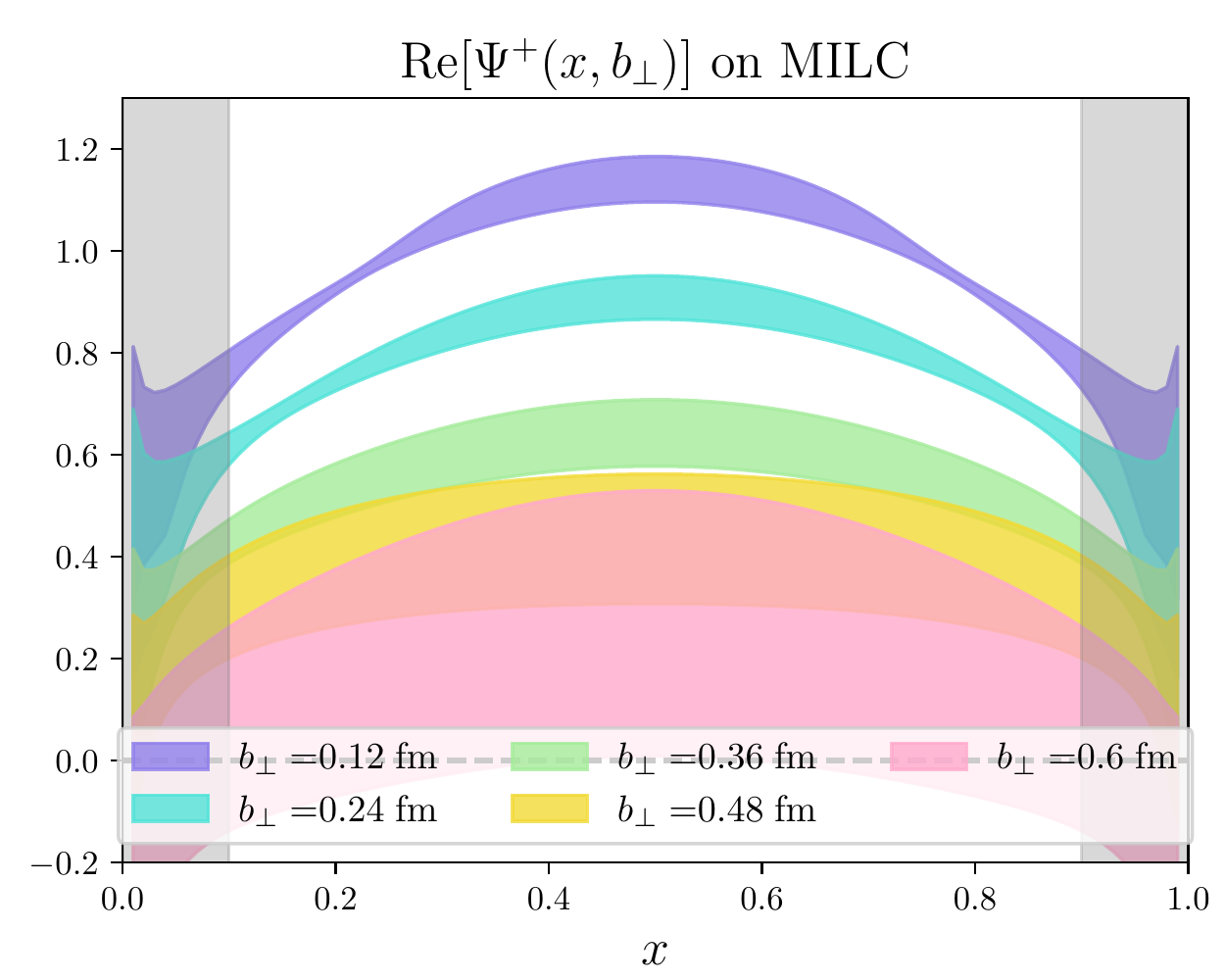}
\includegraphics[scale=0.46]{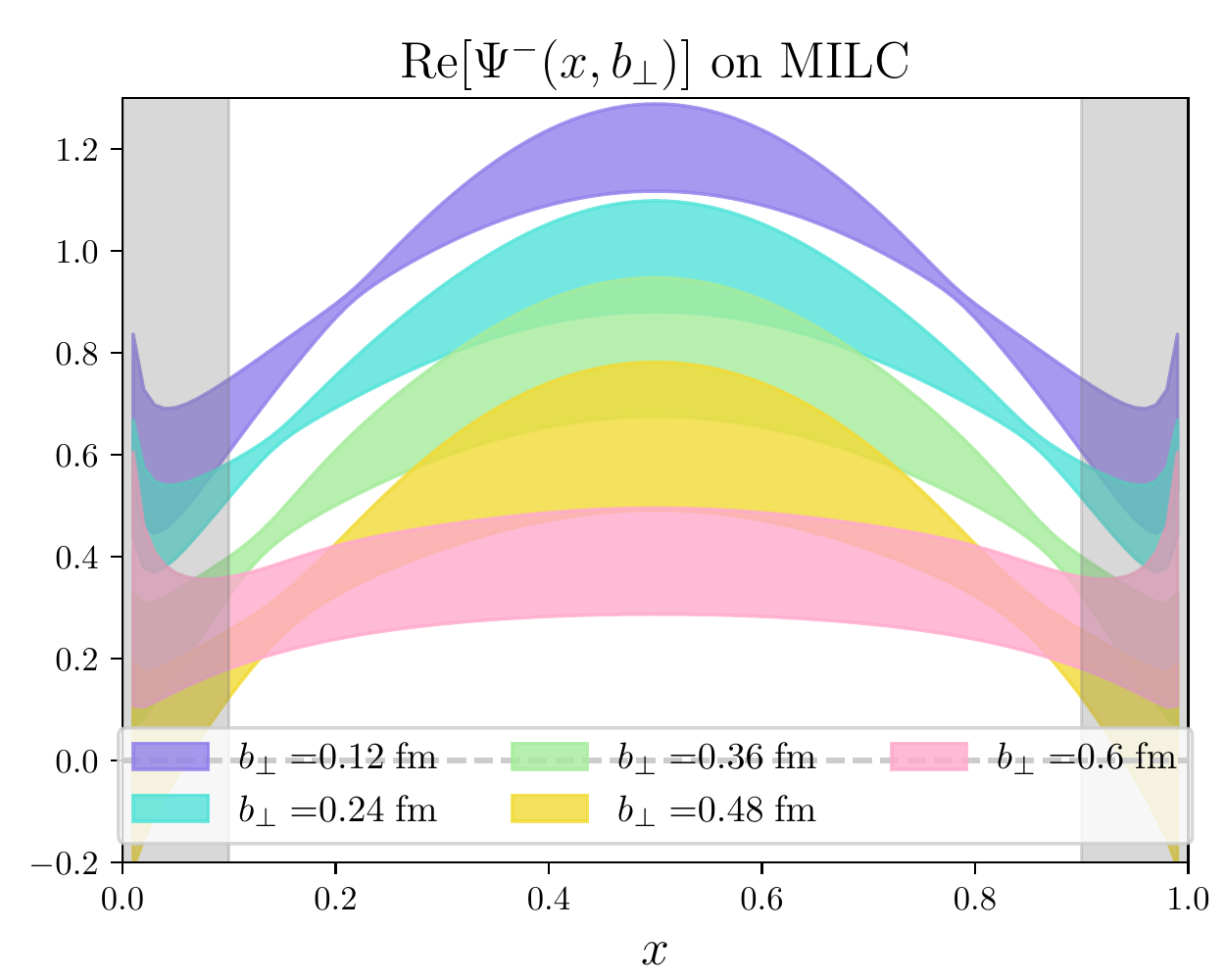}
\includegraphics[scale=0.46]{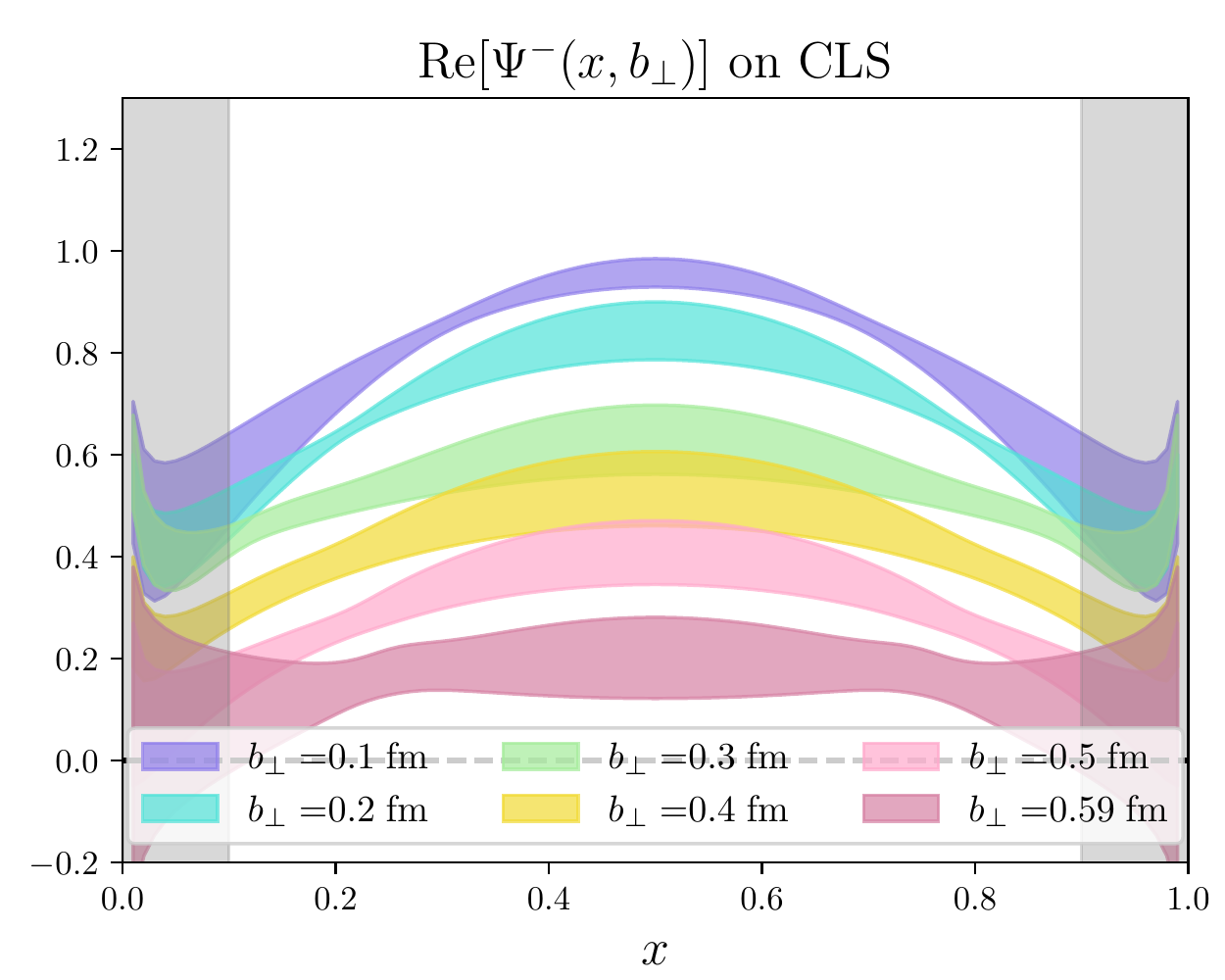}
\includegraphics[scale=0.46]{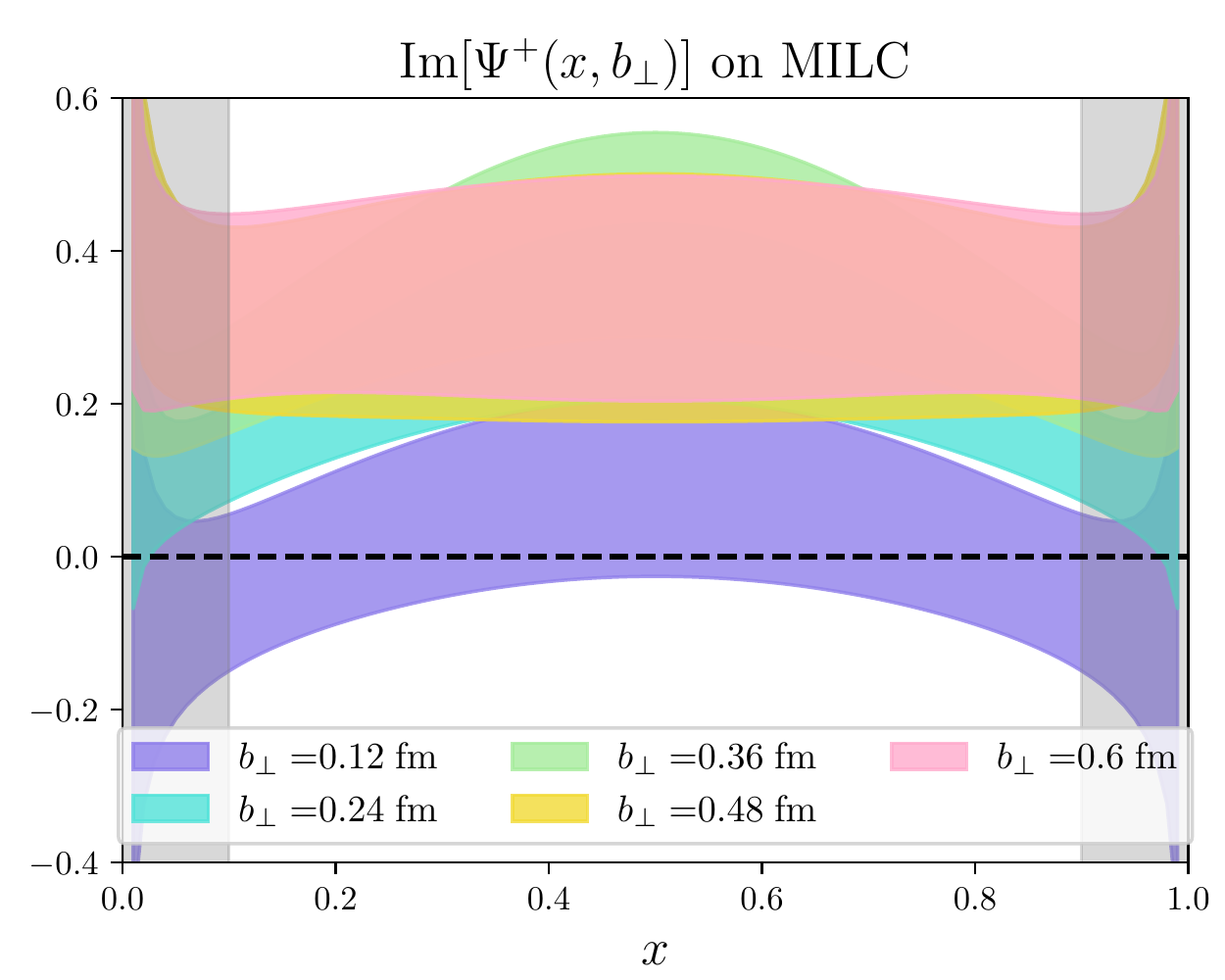}
\includegraphics[scale=0.46]{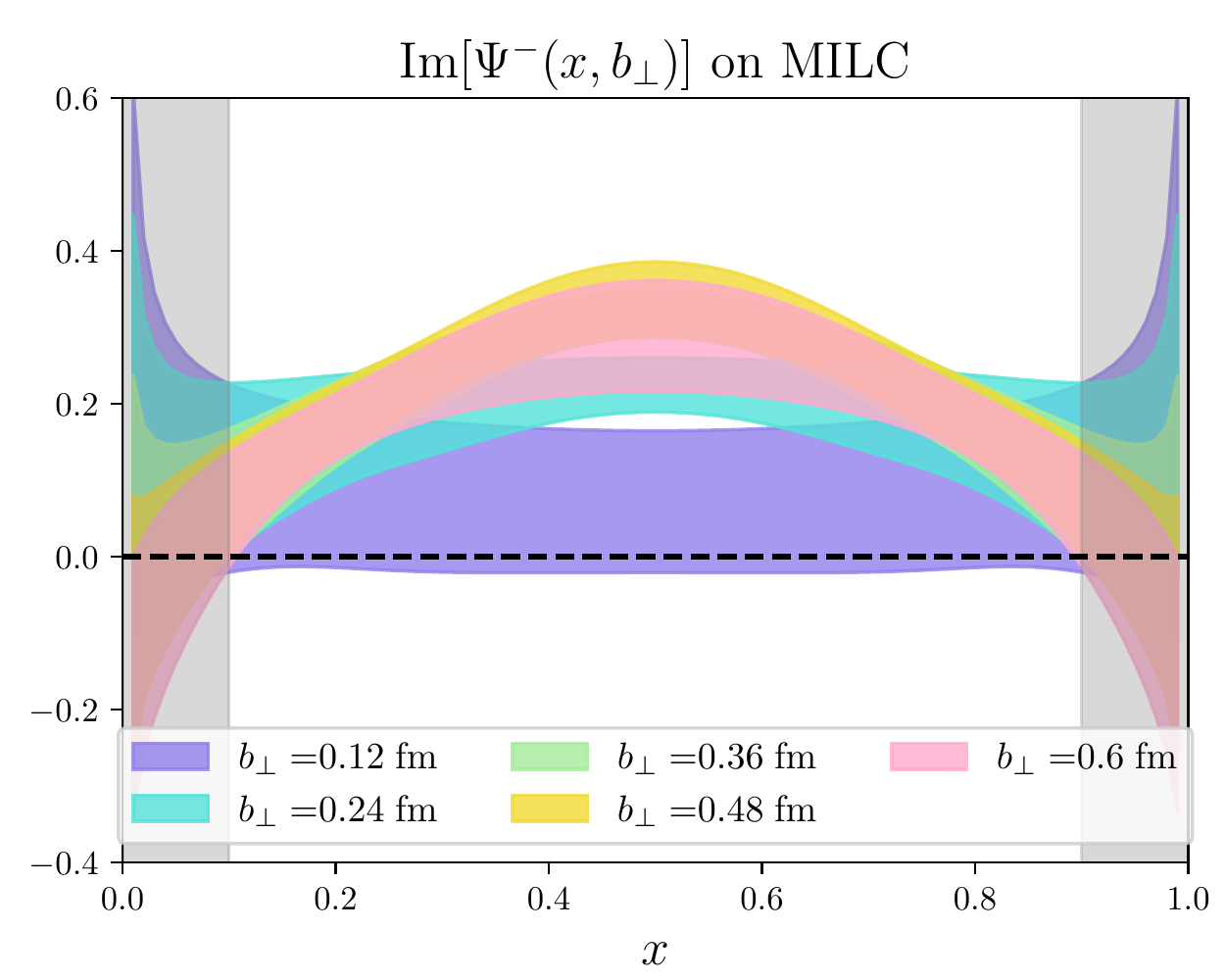}
\includegraphics[scale=0.46]{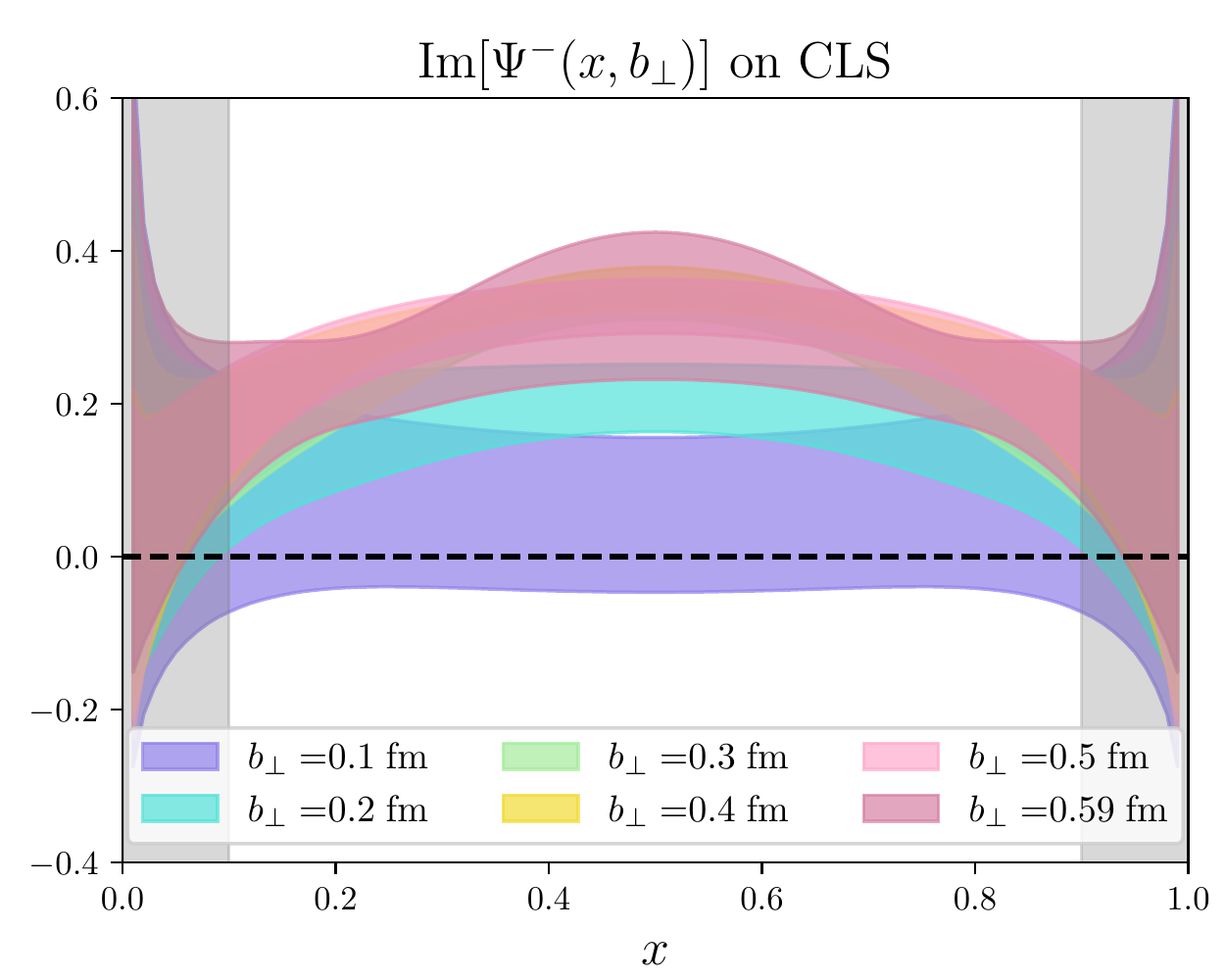}
\caption{The left two figures are for real (upper panel) and imaginary parts (lower panel) of the TMDWF $\Psi^+$ on MILC ensemble, and the central two correspond to $\Psi^-$ on MILC ensemble. The right two ones correspond to $\Psi^-$ on CLS ensemble. These results approach the infinite $P^z$ limit with $\zeta=(6\;\mathrm{GeV})^2$ and $\mu=2$~GeV.}
\label{fig:lcwf_mix}
\end{figure}

\end{widetext}

\bibliography{ref}

\begin{thebibliography}{34}%
\makeatletter
\providecommand \@ifxundefined [1]{%
 \@ifx{#1\undefined}
}%
\providecommand \@ifnum [1]{%
 \ifnum #1\expandafter \@firstoftwo
 \else \expandafter \@secondoftwo
 \fi
}%
\providecommand \@ifx [1]{%
 \ifx #1\expandafter \@firstoftwo
 \else \expandafter \@secondoftwo
 \fi
}%
\providecommand \natexlab [1]{#1}%
\providecommand \enquote  [1]{``#1''}%
\providecommand \bibnamefont  [1]{#1}%
\providecommand \bibfnamefont [1]{#1}%
\providecommand \citenamefont [1]{#1}%
\providecommand \href@noop [0]{\@secondoftwo}%
\providecommand \href [0]{\begingroup \@sanitize@url \@href}%
\providecommand \@href[1]{\@@startlink{#1}\@@href}%
\providecommand \@@href[1]{\endgroup#1\@@endlink}%
\providecommand \@sanitize@url [0]{\catcode `\\12\catcode `\$12\catcode
  `\&12\catcode `\#12\catcode `\^12\catcode `\_12\catcode `\%12\relax}%
\providecommand \@@startlink[1]{}%
\providecommand \@@endlink[0]{}%
\providecommand \url  [0]{\begingroup\@sanitize@url \@url }%
\providecommand \@url [1]{\endgroup\@href {#1}{\urlprefix }}%
\providecommand \urlprefix  [0]{URL }%
\providecommand \Eprint [0]{\href }%
\providecommand \doibase [0]{http://dx.doi.org/}%
\providecommand \selectlanguage [0]{\@gobble}%
\providecommand \bibinfo  [0]{\@secondoftwo}%
\providecommand \bibfield  [0]{\@secondoftwo}%
\providecommand \translation [1]{[#1]}%
\providecommand \BibitemOpen [0]{}%
\providecommand \bibitemStop [0]{}%
\providecommand \bibitemNoStop [0]{.\EOS\space}%
\providecommand \EOS [0]{\spacefactor3000\relax}%
\providecommand \BibitemShut  [1]{\csname bibitem#1\endcsname}%
\let\auto@bib@innerbib\@empty
\bibitem [{\citenamefont {Buchalla}\ \emph {et~al.}(1996)\citenamefont
  {Buchalla}, \citenamefont {Buras},\ and\ \citenamefont
  {Lautenbacher}}]{Buchalla:1995vs}%
  \BibitemOpen
  \bibfield  {author} {\bibinfo {author} {\bibfnamefont {G.}~\bibnamefont
  {Buchalla}}, \bibinfo {author} {\bibfnamefont {A.~J.}\ \bibnamefont {Buras}},
  \ and\ \bibinfo {author} {\bibfnamefont {M.~E.}\ \bibnamefont
  {Lautenbacher}},\ }\href {\doibase 10.1103/RevModPhys.68.1125} {\bibfield
  {journal} {\bibinfo  {journal} {Rev. Mod. Phys.}\ }\textbf {\bibinfo {volume}
  {68}},\ \bibinfo {pages} {1125} (\bibinfo {year} {1996})},\ \Eprint
  {http://arxiv.org/abs/hep-ph/9512380} {arXiv:hep-ph/9512380} \BibitemShut
  {NoStop}%
\bibitem [{\citenamefont {Brodsky}(2001)}]{Brodsky:2001wx}%
  \BibitemOpen
  \bibfield  {author} {\bibinfo {author} {\bibfnamefont {S.~J.}\ \bibnamefont
  {Brodsky}},\ }\href@noop {} {\bibfield  {journal} {\bibinfo  {journal} {Acta
  Phys. Polon. B}\ }\textbf {\bibinfo {volume} {32}},\ \bibinfo {pages} {4013}
  (\bibinfo {year} {2001})},\ \Eprint {http://arxiv.org/abs/hep-ph/0111340}
  {arXiv:hep-ph/0111340} \BibitemShut {NoStop}%
\bibitem [{\citenamefont {Burkardt}\ \emph {et~al.}(2002)\citenamefont
  {Burkardt}, \citenamefont {Ji},\ and\ \citenamefont
  {Yuan}}]{Burkardt:2002uc}%
  \BibitemOpen
  \bibfield  {author} {\bibinfo {author} {\bibfnamefont {M.}~\bibnamefont
  {Burkardt}}, \bibinfo {author} {\bibfnamefont {X.-d.}\ \bibnamefont {Ji}}, \
  and\ \bibinfo {author} {\bibfnamefont {F.}~\bibnamefont {Yuan}},\ }\href
  {\doibase 10.1016/S0370-2693(02)02596-0} {\bibfield  {journal} {\bibinfo
  {journal} {Phys. Lett. B}\ }\textbf {\bibinfo {volume} {545}},\ \bibinfo
  {pages} {345} (\bibinfo {year} {2002})},\ \Eprint
  {http://arxiv.org/abs/hep-ph/0205272} {arXiv:hep-ph/0205272} \BibitemShut
  {NoStop}%
\bibitem [{\citenamefont {Ma}\ and\ \citenamefont {Wang}(2005)}]{Ma:2004ay}%
  \BibitemOpen
  \bibfield  {author} {\bibinfo {author} {\bibfnamefont {J.~P.}\ \bibnamefont
  {Ma}}\ and\ \bibinfo {author} {\bibfnamefont {Q.}~\bibnamefont {Wang}},\
  }\href {\doibase 10.1016/j.physletb.2005.03.049} {\bibfield  {journal}
  {\bibinfo  {journal} {Phys. Lett. B}\ }\textbf {\bibinfo {volume} {613}},\
  \bibinfo {pages} {39} (\bibinfo {year} {2005})},\ \Eprint
  {http://arxiv.org/abs/hep-ph/0412282} {arXiv:hep-ph/0412282} \BibitemShut
  {NoStop}%
\bibitem [{\citenamefont {Collins}\ and\ \citenamefont
  {Soper}(1982)}]{Collins:1981uw}%
  \BibitemOpen
  \bibfield  {author} {\bibinfo {author} {\bibfnamefont {J.~C.}\ \bibnamefont
  {Collins}}\ and\ \bibinfo {author} {\bibfnamefont {D.~E.}\ \bibnamefont
  {Soper}},\ }\href {\doibase 10.1016/0550-3213(82)90021-9} {\bibfield
  {journal} {\bibinfo  {journal} {Nucl. Phys. B}\ }\textbf {\bibinfo {volume}
  {194}},\ \bibinfo {pages} {445} (\bibinfo {year} {1982})}\BibitemShut
  {NoStop}%
\bibitem [{\citenamefont {Keum}\ \emph {et~al.}(2001)\citenamefont {Keum},
  \citenamefont {Li},\ and\ \citenamefont {Sanda}}]{Keum:2000ph}%
  \BibitemOpen
  \bibfield  {author} {\bibinfo {author} {\bibfnamefont {Y.-Y.}\ \bibnamefont
  {Keum}}, \bibinfo {author} {\bibfnamefont {H.-n.}\ \bibnamefont {Li}}, \ and\
  \bibinfo {author} {\bibfnamefont {A.~I.}\ \bibnamefont {Sanda}},\ }\href
  {\doibase 10.1016/S0370-2693(01)00247-7} {\bibfield  {journal} {\bibinfo
  {journal} {Phys. Lett. B}\ }\textbf {\bibinfo {volume} {504}},\ \bibinfo
  {pages} {6} (\bibinfo {year} {2001})},\ \Eprint
  {http://arxiv.org/abs/hep-ph/0004004} {arXiv:hep-ph/0004004} \BibitemShut
  {NoStop}%
\bibitem [{\citenamefont {Lu}\ \emph {et~al.}(2001)\citenamefont {Lu},
  \citenamefont {Ukai},\ and\ \citenamefont {Yang}}]{Lu:2000em}%
  \BibitemOpen
  \bibfield  {author} {\bibinfo {author} {\bibfnamefont {C.-D.}\ \bibnamefont
  {Lu}}, \bibinfo {author} {\bibfnamefont {K.}~\bibnamefont {Ukai}}, \ and\
  \bibinfo {author} {\bibfnamefont {M.-Z.}\ \bibnamefont {Yang}},\ }\href
  {\doibase 10.1103/PhysRevD.63.074009} {\bibfield  {journal} {\bibinfo
  {journal} {Phys. Rev. D}\ }\textbf {\bibinfo {volume} {63}},\ \bibinfo
  {pages} {074009} (\bibinfo {year} {2001})},\ \Eprint
  {http://arxiv.org/abs/hep-ph/0004213} {arXiv:hep-ph/0004213} \BibitemShut
  {NoStop}%
\bibitem [{\citenamefont {Ali}\ \emph {et~al.}(2007)\citenamefont {Ali},
  \citenamefont {Kramer}, \citenamefont {Li}, \citenamefont {Lu}, \citenamefont
  {Shen}, \citenamefont {Wang},\ and\ \citenamefont {Wang}}]{Ali:2007ff}%
  \BibitemOpen
  \bibfield  {author} {\bibinfo {author} {\bibfnamefont {A.}~\bibnamefont
  {Ali}}, \bibinfo {author} {\bibfnamefont {G.}~\bibnamefont {Kramer}},
  \bibinfo {author} {\bibfnamefont {Y.}~\bibnamefont {Li}}, \bibinfo {author}
  {\bibfnamefont {C.-D.}\ \bibnamefont {Lu}}, \bibinfo {author} {\bibfnamefont
  {Y.-L.}\ \bibnamefont {Shen}}, \bibinfo {author} {\bibfnamefont
  {W.}~\bibnamefont {Wang}}, \ and\ \bibinfo {author} {\bibfnamefont {Y.-M.}\
  \bibnamefont {Wang}},\ }\href {\doibase 10.1103/PhysRevD.76.074018}
  {\bibfield  {journal} {\bibinfo  {journal} {Phys. Rev. D}\ }\textbf {\bibinfo
  {volume} {76}},\ \bibinfo {pages} {074018} (\bibinfo {year} {2007})},\
  \Eprint {http://arxiv.org/abs/hep-ph/0703162} {arXiv:hep-ph/0703162}
  \BibitemShut {NoStop}%
\bibitem [{\citenamefont {Ji}(2013)}]{Ji:2013dva}%
  \BibitemOpen
  \bibfield  {author} {\bibinfo {author} {\bibfnamefont {X.}~\bibnamefont
  {Ji}},\ }\href {\doibase 10.1103/PhysRevLett.110.262002} {\bibfield
  {journal} {\bibinfo  {journal} {Phys. Rev. Lett.}\ }\textbf {\bibinfo
  {volume} {110}},\ \bibinfo {pages} {262002} (\bibinfo {year} {2013})},\
  \Eprint {http://arxiv.org/abs/1305.1539} {arXiv:1305.1539 [hep-ph]}
  \BibitemShut {NoStop}%
\bibitem [{\citenamefont {Ji}\ \emph {et~al.}(2015)\citenamefont {Ji},
  \citenamefont {Sun}, \citenamefont {Xiong},\ and\ \citenamefont
  {Yuan}}]{Ji:2014hxa}%
  \BibitemOpen
  \bibfield  {author} {\bibinfo {author} {\bibfnamefont {X.}~\bibnamefont
  {Ji}}, \bibinfo {author} {\bibfnamefont {P.}~\bibnamefont {Sun}}, \bibinfo
  {author} {\bibfnamefont {X.}~\bibnamefont {Xiong}}, \ and\ \bibinfo {author}
  {\bibfnamefont {F.}~\bibnamefont {Yuan}},\ }\href {\doibase
  10.1103/PhysRevD.91.074009} {\bibfield  {journal} {\bibinfo  {journal} {Phys.
  Rev. D}\ }\textbf {\bibinfo {volume} {91}},\ \bibinfo {pages} {074009}
  (\bibinfo {year} {2015})},\ \Eprint {http://arxiv.org/abs/1405.7640}
  {arXiv:1405.7640 [hep-ph]} \BibitemShut {NoStop}%
\bibitem [{\citenamefont {Ji}\ \emph {et~al.}(2021{\natexlab{a}})\citenamefont
  {Ji}, \citenamefont {Liu}, \citenamefont {Liu}, \citenamefont {Zhang},\ and\
  \citenamefont {Zhao}}]{Ji:2020ect}%
  \BibitemOpen
  \bibfield  {author} {\bibinfo {author} {\bibfnamefont {X.}~\bibnamefont
  {Ji}}, \bibinfo {author} {\bibfnamefont {Y.-S.}\ \bibnamefont {Liu}},
  \bibinfo {author} {\bibfnamefont {Y.}~\bibnamefont {Liu}}, \bibinfo {author}
  {\bibfnamefont {J.-H.}\ \bibnamefont {Zhang}}, \ and\ \bibinfo {author}
  {\bibfnamefont {Y.}~\bibnamefont {Zhao}},\ }\href {\doibase
  10.1103/RevModPhys.93.035005} {\bibfield  {journal} {\bibinfo  {journal}
  {Rev. Mod. Phys.}\ }\textbf {\bibinfo {volume} {93}},\ \bibinfo {pages}
  {035005} (\bibinfo {year} {2021}{\natexlab{a}})},\ \Eprint
  {http://arxiv.org/abs/2004.03543} {arXiv:2004.03543 [hep-ph]} \BibitemShut
  {NoStop}%
\bibitem [{\citenamefont {Collins}\ and\ \citenamefont
  {Soper}(1981)}]{Collins:1981uk}%
  \BibitemOpen
  \bibfield  {author} {\bibinfo {author} {\bibfnamefont {J.~C.}\ \bibnamefont
  {Collins}}\ and\ \bibinfo {author} {\bibfnamefont {D.~E.}\ \bibnamefont
  {Soper}},\ }\href {\doibase 10.1016/0550-3213(81)90339-4} {\bibfield
  {journal} {\bibinfo  {journal} {Nucl. Phys. B}\ }\textbf {\bibinfo {volume}
  {193}},\ \bibinfo {pages} {381} (\bibinfo {year} {1981})},\ \bibinfo {note}
  {[Erratum: Nucl.Phys.B 213, 545 (1983)]}\BibitemShut {NoStop}%
\bibitem [{\citenamefont {Collins}\ \emph {et~al.}(1988)\citenamefont
  {Collins}, \citenamefont {Soper},\ and\ \citenamefont
  {Sterman}}]{Collins:1988ig}%
  \BibitemOpen
  \bibfield  {author} {\bibinfo {author} {\bibfnamefont {J.~C.}\ \bibnamefont
  {Collins}}, \bibinfo {author} {\bibfnamefont {D.~E.}\ \bibnamefont {Soper}},
  \ and\ \bibinfo {author} {\bibfnamefont {G.~F.}\ \bibnamefont {Sterman}},\
  }\href {\doibase 10.1016/0550-3213(88)90130-7} {\bibfield  {journal}
  {\bibinfo  {journal} {Nucl. Phys. B}\ }\textbf {\bibinfo {volume} {308}},\
  \bibinfo {pages} {833} (\bibinfo {year} {1988})}\BibitemShut {NoStop}%
\bibitem [{\citenamefont {Ji}\ \emph {et~al.}(2020{\natexlab{a}})\citenamefont
  {Ji}, \citenamefont {Liu},\ and\ \citenamefont {Liu}}]{Ji:2019ewn}%
  \BibitemOpen
  \bibfield  {author} {\bibinfo {author} {\bibfnamefont {X.}~\bibnamefont
  {Ji}}, \bibinfo {author} {\bibfnamefont {Y.}~\bibnamefont {Liu}}, \ and\
  \bibinfo {author} {\bibfnamefont {Y.-S.}\ \bibnamefont {Liu}},\ }\href
  {\doibase 10.1016/j.physletb.2020.135946} {\bibfield  {journal} {\bibinfo
  {journal} {Phys. Lett. B}\ }\textbf {\bibinfo {volume} {811}},\ \bibinfo
  {pages} {135946} (\bibinfo {year} {2020}{\natexlab{a}})},\ \Eprint
  {http://arxiv.org/abs/1911.03840} {arXiv:1911.03840 [hep-ph]} \BibitemShut
  {NoStop}%
\bibitem [{\citenamefont {Ji}\ \emph {et~al.}(2020{\natexlab{b}})\citenamefont
  {Ji}, \citenamefont {Liu},\ and\ \citenamefont {Liu}}]{Ji:2019sxk}%
  \BibitemOpen
  \bibfield  {author} {\bibinfo {author} {\bibfnamefont {X.}~\bibnamefont
  {Ji}}, \bibinfo {author} {\bibfnamefont {Y.}~\bibnamefont {Liu}}, \ and\
  \bibinfo {author} {\bibfnamefont {Y.-S.}\ \bibnamefont {Liu}},\ }\href
  {\doibase 10.1016/j.nuclphysb.2020.115054} {\bibfield  {journal} {\bibinfo
  {journal} {Nucl. Phys. B}\ }\textbf {\bibinfo {volume} {955}},\ \bibinfo
  {pages} {115054} (\bibinfo {year} {2020}{\natexlab{b}})},\ \Eprint
  {http://arxiv.org/abs/1910.11415} {arXiv:1910.11415 [hep-ph]} \BibitemShut
  {NoStop}%
\bibitem [{\citenamefont {Ji}\ and\ \citenamefont {Liu}(2022)}]{Ji:2021znw}%
  \BibitemOpen
  \bibfield  {author} {\bibinfo {author} {\bibfnamefont {X.}~\bibnamefont
  {Ji}}\ and\ \bibinfo {author} {\bibfnamefont {Y.}~\bibnamefont {Liu}},\
  }\href {\doibase 10.1103/PhysRevD.105.076014} {\bibfield  {journal} {\bibinfo
   {journal} {Phys. Rev. D}\ }\textbf {\bibinfo {volume} {105}},\ \bibinfo
  {pages} {076014} (\bibinfo {year} {2022})},\ \Eprint
  {http://arxiv.org/abs/2106.05310} {arXiv:2106.05310 [hep-ph]} \BibitemShut
  {NoStop}%
\bibitem [{\citenamefont {Shanahan}\ \emph {et~al.}(2021)\citenamefont
  {Shanahan}, \citenamefont {Wagman},\ and\ \citenamefont
  {Zhao}}]{Shanahan:2021tst}%
  \BibitemOpen
  \bibfield  {author} {\bibinfo {author} {\bibfnamefont {P.}~\bibnamefont
  {Shanahan}}, \bibinfo {author} {\bibfnamefont {M.}~\bibnamefont {Wagman}}, \
  and\ \bibinfo {author} {\bibfnamefont {Y.}~\bibnamefont {Zhao}},\ }\href
  {\doibase 10.1103/PhysRevD.104.114502} {\bibfield  {journal} {\bibinfo
  {journal} {Phys. Rev. D}\ }\textbf {\bibinfo {volume} {104}},\ \bibinfo
  {pages} {114502} (\bibinfo {year} {2021})},\ \Eprint
  {http://arxiv.org/abs/2107.11930} {arXiv:2107.11930 [hep-lat]} \BibitemShut
  {NoStop}%
\bibitem [{\citenamefont {Chu}\ \emph {et~al.}(2022)\citenamefont {Chu} \emph
  {et~al.}}]{LPC:2022ibr}%
  \BibitemOpen
  \bibfield  {author} {\bibinfo {author} {\bibfnamefont {M.-H.}\ \bibnamefont
  {Chu}} \emph {et~al.} (\bibinfo {collaboration} {LPC}),\ }\href {\doibase
  10.1103/PhysRevD.106.034509} {\bibfield  {journal} {\bibinfo  {journal}
  {Phys. Rev. D}\ }\textbf {\bibinfo {volume} {106}},\ \bibinfo {pages}
  {034509} (\bibinfo {year} {2022})},\ \Eprint
  {http://arxiv.org/abs/2204.00200} {arXiv:2204.00200 [hep-lat]} \BibitemShut
  {NoStop}%
\bibitem [{\citenamefont {Schlemmer}\ \emph {et~al.}(2021)\citenamefont
  {Schlemmer}, \citenamefont {Vladimirov}, \citenamefont {Zimmermann},
  \citenamefont {Engelhardt},\ and\ \citenamefont
  {Sch\"afer}}]{Schlemmer:2021aij}%
  \BibitemOpen
  \bibfield  {author} {\bibinfo {author} {\bibfnamefont {M.}~\bibnamefont
  {Schlemmer}}, \bibinfo {author} {\bibfnamefont {A.}~\bibnamefont
  {Vladimirov}}, \bibinfo {author} {\bibfnamefont {C.}~\bibnamefont
  {Zimmermann}}, \bibinfo {author} {\bibfnamefont {M.}~\bibnamefont
  {Engelhardt}}, \ and\ \bibinfo {author} {\bibfnamefont {A.}~\bibnamefont
  {Sch\"afer}},\ }\href {\doibase 10.1007/JHEP08(2021)004} {\bibfield
  {journal} {\bibinfo  {journal} {JHEP}\ }\textbf {\bibinfo {volume} {08}},\
  \bibinfo {pages} {004} (\bibinfo {year} {2021})},\ \Eprint
  {http://arxiv.org/abs/2103.16991} {arXiv:2103.16991 [hep-lat]} \BibitemShut
  {NoStop}%
\bibitem [{\citenamefont {Deng}\ \emph {et~al.}(2022)\citenamefont {Deng},
  \citenamefont {Wang},\ and\ \citenamefont {Zeng}}]{Deng:2022gzi}%
  \BibitemOpen
  \bibfield  {author} {\bibinfo {author} {\bibfnamefont {Z.-F.}\ \bibnamefont
  {Deng}}, \bibinfo {author} {\bibfnamefont {W.}~\bibnamefont {Wang}}, \ and\
  \bibinfo {author} {\bibfnamefont {J.}~\bibnamefont {Zeng}},\ }\href {\doibase
  10.1007/JHEP09(2022)046} {\bibfield  {journal} {\bibinfo  {journal} {JHEP}\
  }\textbf {\bibinfo {volume} {09}},\ \bibinfo {pages} {046} (\bibinfo {year}
  {2022})},\ \Eprint {http://arxiv.org/abs/2207.07280} {arXiv:2207.07280
  [hep-th]} \BibitemShut {NoStop}%
\bibitem [{\citenamefont {Zhang}\ \emph {et~al.}(2022)\citenamefont {Zhang},
  \citenamefont {Ji}, \citenamefont {Yang}, \citenamefont {Yao},\ and\
  \citenamefont {Zhang}}]{Zhang:2022xuw}%
  \BibitemOpen
  \bibfield  {author} {\bibinfo {author} {\bibfnamefont {K.}~\bibnamefont
  {Zhang}}, \bibinfo {author} {\bibfnamefont {X.}~\bibnamefont {Ji}}, \bibinfo
  {author} {\bibfnamefont {Y.-B.}\ \bibnamefont {Yang}}, \bibinfo {author}
  {\bibfnamefont {F.}~\bibnamefont {Yao}}, \ and\ \bibinfo {author}
  {\bibfnamefont {J.-H.}\ \bibnamefont {Zhang}} (\bibinfo {collaboration}
  {[Lattice Parton Collaboration (LPC)]}),\ }\href {\doibase
  10.1103/PhysRevLett.129.082002} {\bibfield  {journal} {\bibinfo  {journal}
  {Phys. Rev. Lett.}\ }\textbf {\bibinfo {volume} {129}},\ \bibinfo {pages}
  {082002} (\bibinfo {year} {2022})},\ \Eprint
  {http://arxiv.org/abs/2205.13402} {arXiv:2205.13402 [hep-lat]} \BibitemShut
  {NoStop}%
\bibitem [{sup()}]{supplemental}%
  \BibitemOpen
  \href@noop {} {\bibinfo  {journal} {Supplemental Material}\ }\BibitemShut
  {NoStop}%
\bibitem [{\citenamefont {Follana}\ \emph {et~al.}(2007)\citenamefont
  {Follana}, \citenamefont {Mason}, \citenamefont {Davies}, \citenamefont
  {Hornbostel}, \citenamefont {Lepage}, \citenamefont {Shigemitsu},
  \citenamefont {Trottier},\ and\ \citenamefont {Wong}}]{Follana:2006rc}%
  \BibitemOpen
\bibfield  {journal} {  }\bibfield  {author} {\bibinfo {author} {\bibfnamefont
  {E.}~\bibnamefont {Follana}}, \bibinfo {author} {\bibfnamefont
  {Q.}~\bibnamefont {Mason}}, \bibinfo {author} {\bibfnamefont
  {C.}~\bibnamefont {Davies}}, \bibinfo {author} {\bibfnamefont
  {K.}~\bibnamefont {Hornbostel}}, \bibinfo {author} {\bibfnamefont {G.~P.}\
  \bibnamefont {Lepage}}, \bibinfo {author} {\bibfnamefont {J.}~\bibnamefont
  {Shigemitsu}}, \bibinfo {author} {\bibfnamefont {H.}~\bibnamefont
  {Trottier}}, \ and\ \bibinfo {author} {\bibfnamefont {K.}~\bibnamefont
  {Wong}} (\bibinfo {collaboration} {HPQCD, UKQCD}),\ }\href {\doibase
  10.1103/PhysRevD.75.054502} {\bibfield  {journal} {\bibinfo  {journal} {Phys.
  Rev. D}\ }\textbf {\bibinfo {volume} {75}},\ \bibinfo {pages} {054502}
  (\bibinfo {year} {2007})},\ \Eprint {http://arxiv.org/abs/hep-lat/0610092}
  {arXiv:hep-lat/0610092} \BibitemShut {NoStop}%
\bibitem [{\citenamefont {Bazavov}\ \emph {et~al.}(2013)\citenamefont {Bazavov}
  \emph {et~al.}}]{MILC:2012znn}%
  \BibitemOpen
  \bibfield  {author} {\bibinfo {author} {\bibfnamefont {A.}~\bibnamefont
  {Bazavov}} \emph {et~al.} (\bibinfo {collaboration} {MILC}),\ }\href
  {\doibase 10.1103/PhysRevD.87.054505} {\bibfield  {journal} {\bibinfo
  {journal} {Phys. Rev. D}\ }\textbf {\bibinfo {volume} {87}},\ \bibinfo
  {pages} {054505} (\bibinfo {year} {2013})},\ \Eprint
  {http://arxiv.org/abs/1212.4768} {arXiv:1212.4768 [hep-lat]} \BibitemShut
  {NoStop}%
\bibitem [{\citenamefont {Hasenfratz}\ and\ \citenamefont
  {Knechtli}(2001)}]{Hasenfratz:2001hp}%
  \BibitemOpen
  \bibfield  {author} {\bibinfo {author} {\bibfnamefont {A.}~\bibnamefont
  {Hasenfratz}}\ and\ \bibinfo {author} {\bibfnamefont {F.}~\bibnamefont
  {Knechtli}},\ }\href {\doibase 10.1103/PhysRevD.64.034504} {\bibfield
  {journal} {\bibinfo  {journal} {Phys. Rev. D}\ }\textbf {\bibinfo {volume}
  {64}},\ \bibinfo {pages} {034504} (\bibinfo {year} {2001})},\ \Eprint
  {http://arxiv.org/abs/hep-lat/0103029} {arXiv:hep-lat/0103029} \BibitemShut
  {NoStop}%
\bibitem [{\citenamefont {Ji}\ \emph {et~al.}(2021{\natexlab{b}})\citenamefont
  {Ji}, \citenamefont {Liu}, \citenamefont {Sch\"afer}, \citenamefont {Wang},
  \citenamefont {Yang}, \citenamefont {Zhang},\ and\ \citenamefont
  {Zhao}}]{Ji:2020brr}%
  \BibitemOpen
  \bibfield  {author} {\bibinfo {author} {\bibfnamefont {X.}~\bibnamefont
  {Ji}}, \bibinfo {author} {\bibfnamefont {Y.}~\bibnamefont {Liu}}, \bibinfo
  {author} {\bibfnamefont {A.}~\bibnamefont {Sch\"afer}}, \bibinfo {author}
  {\bibfnamefont {W.}~\bibnamefont {Wang}}, \bibinfo {author} {\bibfnamefont
  {Y.-B.}\ \bibnamefont {Yang}}, \bibinfo {author} {\bibfnamefont {J.-H.}\
  \bibnamefont {Zhang}}, \ and\ \bibinfo {author} {\bibfnamefont
  {Y.}~\bibnamefont {Zhao}},\ }\href {\doibase 10.1016/j.nuclphysb.2021.115311}
  {\bibfield  {journal} {\bibinfo  {journal} {Nucl. Phys. B}\ }\textbf
  {\bibinfo {volume} {964}},\ \bibinfo {pages} {115311} (\bibinfo {year}
  {2021}{\natexlab{b}})},\ \Eprint {http://arxiv.org/abs/2008.03886}
  {arXiv:2008.03886 [hep-ph]} \BibitemShut {NoStop}%
\bibitem [{\citenamefont {Hua}\ \emph {et~al.}(2022)\citenamefont {Hua} \emph
  {et~al.}}]{LatticeParton:2022zqc}%
  \BibitemOpen
  \bibfield  {author} {\bibinfo {author} {\bibfnamefont {J.}~\bibnamefont
  {Hua}} \emph {et~al.} (\bibinfo {collaboration} {Lattice Parton}),\ }\href
  {\doibase 10.1103/PhysRevLett.129.132001} {\bibfield  {journal} {\bibinfo
  {journal} {Phys. Rev. Lett.}\ }\textbf {\bibinfo {volume} {129}},\ \bibinfo
  {pages} {132001} (\bibinfo {year} {2022})},\ \Eprint
  {http://arxiv.org/abs/2201.09173} {arXiv:2201.09173 [hep-lat]} \BibitemShut
  {NoStop}%
\bibitem [{\citenamefont {Zhang}\ \emph {et~al.}(2020)\citenamefont {Zhang}
  \emph {et~al.}}]{LatticeParton:2020uhz}%
  \BibitemOpen
  \bibfield  {author} {\bibinfo {author} {\bibfnamefont {Q.-A.}\ \bibnamefont
  {Zhang}} \emph {et~al.} (\bibinfo {collaboration} {Lattice Parton}),\ }\href
  {\doibase 10.22323/1.396.0477} {\bibfield  {journal} {\bibinfo  {journal}
  {Phys. Rev. Lett.}\ }\textbf {\bibinfo {volume} {125}},\ \bibinfo {pages}
  {192001} (\bibinfo {year} {2020})},\ \Eprint
  {http://arxiv.org/abs/2005.14572} {arXiv:2005.14572 [hep-lat]} \BibitemShut
  {NoStop}%
\bibitem [{\citenamefont {Li}\ \emph {et~al.}(2022)\citenamefont {Li} \emph
  {et~al.}}]{Li:2021wvl}%
  \BibitemOpen
  \bibfield  {author} {\bibinfo {author} {\bibfnamefont {Y.}~\bibnamefont {Li}}
  \emph {et~al.},\ }\href {\doibase 10.1103/PhysRevLett.128.062002} {\bibfield
  {journal} {\bibinfo  {journal} {Phys. Rev. Lett.}\ }\textbf {\bibinfo
  {volume} {128}},\ \bibinfo {pages} {062002} (\bibinfo {year} {2022})},\
  \Eprint {http://arxiv.org/abs/2106.13027} {arXiv:2106.13027 [hep-lat]}
  \BibitemShut {NoStop}%
\bibitem [{\citenamefont {Ebert}\ \emph {et~al.}(2019)\citenamefont {Ebert},
  \citenamefont {Stewart},\ and\ \citenamefont {Zhao}}]{Ebert:2019okf}%
  \BibitemOpen
  \bibfield  {author} {\bibinfo {author} {\bibfnamefont {M.~A.}\ \bibnamefont
  {Ebert}}, \bibinfo {author} {\bibfnamefont {I.~W.}\ \bibnamefont {Stewart}},
  \ and\ \bibinfo {author} {\bibfnamefont {Y.}~\bibnamefont {Zhao}},\ }\href
  {\doibase 10.1007/JHEP09(2019)037} {\bibfield  {journal} {\bibinfo  {journal}
  {JHEP}\ }\textbf {\bibinfo {volume} {09}},\ \bibinfo {pages} {037} (\bibinfo
  {year} {2019})},\ \Eprint {http://arxiv.org/abs/1901.03685} {arXiv:1901.03685
  [hep-ph]} \BibitemShut {NoStop}%
\bibitem [{\citenamefont {Hua}\ \emph {et~al.}(2021)\citenamefont {Hua},
  \citenamefont {Chu}, \citenamefont {Sun}, \citenamefont {Wang}, \citenamefont
  {Xu}, \citenamefont {Yang}, \citenamefont {Zhang},\ and\ \citenamefont
  {Zhang}}]{Hua:2020gnw}%
  \BibitemOpen
  \bibfield  {author} {\bibinfo {author} {\bibfnamefont {J.}~\bibnamefont
  {Hua}}, \bibinfo {author} {\bibfnamefont {M.-H.}\ \bibnamefont {Chu}},
  \bibinfo {author} {\bibfnamefont {P.}~\bibnamefont {Sun}}, \bibinfo {author}
  {\bibfnamefont {W.}~\bibnamefont {Wang}}, \bibinfo {author} {\bibfnamefont
  {J.}~\bibnamefont {Xu}}, \bibinfo {author} {\bibfnamefont {Y.-B.}\
  \bibnamefont {Yang}}, \bibinfo {author} {\bibfnamefont {J.-H.}\ \bibnamefont
  {Zhang}}, \ and\ \bibinfo {author} {\bibfnamefont {Q.-A.}\ \bibnamefont
  {Zhang}} (\bibinfo {collaboration} {Lattice Parton}),\ }\href {\doibase
  10.1103/PhysRevLett.127.062002} {\bibfield  {journal} {\bibinfo  {journal}
  {Phys. Rev. Lett.}\ }\textbf {\bibinfo {volume} {127}},\ \bibinfo {pages}
  {062002} (\bibinfo {year} {2021})},\ \Eprint
  {http://arxiv.org/abs/2011.09788} {arXiv:2011.09788 [hep-lat]} \BibitemShut
  {NoStop}%
\bibitem [{\citenamefont {Lu}\ \emph {et~al.}(2007)\citenamefont {Lu},
  \citenamefont {Wang},\ and\ \citenamefont {Wang}}]{Lu:2007hr}%
  \BibitemOpen
  \bibfield  {author} {\bibinfo {author} {\bibfnamefont {C.-D.}\ \bibnamefont
  {Lu}}, \bibinfo {author} {\bibfnamefont {W.}~\bibnamefont {Wang}}, \ and\
  \bibinfo {author} {\bibfnamefont {Y.-M.}\ \bibnamefont {Wang}},\ }\href
  {\doibase 10.1103/PhysRevD.75.094020} {\bibfield  {journal} {\bibinfo
  {journal} {Phys. Rev. D}\ }\textbf {\bibinfo {volume} {75}},\ \bibinfo
  {pages} {094020} (\bibinfo {year} {2007})},\ \Eprint
  {http://arxiv.org/abs/hep-ph/0702085} {arXiv:hep-ph/0702085} \BibitemShut
  {NoStop}%
\bibitem [{\citenamefont {Ji}\ \emph {et~al.}(2018)\citenamefont {Ji},
  \citenamefont {Zhang},\ and\ \citenamefont {Zhao}}]{Ji:2017oey}%
  \BibitemOpen
  \bibfield  {author} {\bibinfo {author} {\bibfnamefont {X.}~\bibnamefont
  {Ji}}, \bibinfo {author} {\bibfnamefont {J.-H.}\ \bibnamefont {Zhang}}, \
  and\ \bibinfo {author} {\bibfnamefont {Y.}~\bibnamefont {Zhao}},\ }\href
  {\doibase 10.1103/PhysRevLett.120.112001} {\bibfield  {journal} {\bibinfo
  {journal} {Phys. Rev. Lett.}\ }\textbf {\bibinfo {volume} {120}},\ \bibinfo
  {pages} {112001} (\bibinfo {year} {2018})},\ \Eprint
  {http://arxiv.org/abs/1706.08962} {arXiv:1706.08962 [hep-ph]} \BibitemShut
  {NoStop}%
\bibitem [{\citenamefont {Huo}\ \emph {et~al.}(2021)\citenamefont {Huo} \emph
  {et~al.}}]{LatticePartonCollaborationLPC:2021xdx}%
  \BibitemOpen
  \bibfield  {author} {\bibinfo {author} {\bibfnamefont {Y.-K.}\ \bibnamefont
  {Huo}} \emph {et~al.} (\bibinfo {collaboration} {Lattice Parton Collaboration
  (LPC)}),\ }\href {\doibase 10.1016/j.nuclphysb.2021.115443} {\bibfield
  {journal} {\bibinfo  {journal} {Nucl. Phys. B}\ }\textbf {\bibinfo {volume}
  {969}},\ \bibinfo {pages} {115443} (\bibinfo {year} {2021})},\ \Eprint
  {http://arxiv.org/abs/2103.02965} {arXiv:2103.02965 [hep-lat]} \BibitemShut
  {NoStop}%
\end{thebibliography}%


\end{document}